\documentclass[apj]{emulateapj}
\usepackage{amsmath}
\usepackage{graphicx}
\usepackage{amssymb}
\usepackage{times}
\usepackage{xcolor}
\usepackage{orcidlink}
\newcommand{\AuthorORCID}[2]{%
  \href{https://orcid.org/#2}{\mbox{#1~{\Large\orcidlink{#2}}\kern-0.5em}}%
  }

\usepackage[normalem]{ulem}

\usepackage{etoolbox}
\makeatletter
\patchcmd{\acknowledgments}
  {\begin{internallinenumbers}}
  {\ifnumlines\begin{internallinenumbers}\fi}
  {}
  {\GenericWarning{}{Could not patch acknowledgments start}}
\patchcmd{\endacknowledgments}
  {\end{internallinenumbers}}
  {\ifnumlines\end{internallinenumbers}\fi}
  {}
  {\GenericWarning{}{Could not patch acknowledgments end}}
\makeatother

\begin{document}

\title{FRB\,121102: No supernova-like ejecta or magnetar power, hinting at a binary WD merger}

\shorttitle{FRB121102: No supernova ejecta or magnetar}
\shortauthors{Waxman, Ofek \& Kushnir}

\author{\AuthorORCID{Eli Waxman}{0000-0002-9038-5877}, \AuthorORCID{Eran O. Ofek}{0000-0002-6786-8774}, \AuthorORCID{Doron Kushnir}{0009-0001-3821-8318}}
\affiliation{Dept. of Particle Phys. \& Astrophys., Weizmann Institute of Science, Rehovot 76100, Israel}

\begin{abstract}

We use measurements of the time-dependent dispersion measure of the repeating FRB~121102, together with earlier radio observations of its associated persistent radio source (PRS), to derive stringent constraints on its underlying ``engine.'' The energy held by the relativistic PRS plasma is $>10^{49.5}$\,erg, and its age is $\approx60$\,yr, corresponding to an underlying source power exceeding $10^{40}$\,erg\,s$^{-1}$. If the underlying source is a neutron star, this implies a rotational rather than a magnetic energy source, consistent with a $\sim10$\,ms, $\sim10^{12.5}$\,G neutron star. Alternatively, accretion may also be a viable energy source. The velocity and the kinetic energy of the cold plasma confining the relativistic PRS plasma are inconsistent with it being typical supernova ejecta (unless a significant fraction of the ejecta mass is carried by high-density clumps of $\approx10^{-2.5}$ fractional size)- its expansion speed is limited to a few hundred km/s, which also implies that it was ejected from the source more than $\approx 10^3$~yr preceding the onset of relativistic PRS plasma emission. These constraints may be satisfied by a white dwarf binary merger progenitor system, where a fraction of a solar mass was ejected at a slow speed during and after the merger, and a rapidly rotating neutron star was formed after $\sim10^3$~yr thermal evolution period of the merger remnant.
\end{abstract}

\section{Introduction}
\label{sec:intro}

The sources and mechanisms producing fast radio bursts (FRBs), bright ($0.1-10$\,Jy) millisecond bursts of radio waves in the GHz range \citep{Lorimer+2007Sci_FRB_Discovery}, remain unknown \citep[see][for reviews]{Katz16Rev,2019Cordes-FRBReview,2019Petroff-FRBrev,2024Lorimer--FRBrev,2024Zhang-FRBrev}. Persistent radio emission has been identified in association with a handful of repeating FRBs (see Table~\ref{tab:PRSs}). Observed properties of the persistent radio sources (PRSs), which are likely very rare objects \citep[less than $10^{-3}$ per Milky Way mass galaxy,][]{Ofek2017_FRB121102_like_Search}, allow constraints to be placed on the sources and their environments, which cannot otherwise be obtained from the measured FRB properties alone. This is a great advantage. However, since only a handful of PRSs were identified in association with FRBs, the conclusions may not apply to the entire FRB population.

In this paper, we return to the analysis of the properties of the repeating FRB\,20121102A \citep[hereafter FRB\,121102,][]{Spitler16Repeat,Scholz16Repeating}, for which sub-arc-second localization \citep{Chatterjee17FRB_loc} led to the identification of a dwarf galaxy at a redshift of $z=0.19$ \citep{2017ApJ...834L...7T} and of a PRS \citep{Chatterjee17FRB_loc}, both located in a direction consistent with that of the FRB. The measurements of a rise followed by a decline of the dispersion measure (DM) of the bursts \citep[][]{2014Spitler-121102DM,Spitler16Repeat,2019Hessels-121102DM,2020Oostrum-121102DM,2020Majid-121102DM,2021Li-121102DM,2021Platts-121102DM,2023Snelders-121102DM,2025Wang-121102DM,2025Zhang-121102DM,2025Snelders-FRB121102} enable us to draw more detailed constraints on the source than those obtained in our earlier analysis \citep[][]{2017Waxman-FRB}, which was carried out at a time when only upper limits were available on the DM variations.

PRS models typically attribute the persistent radio emission to a relativistic plasma ``bubble'' confined by a colder, denser plasma shell. The most widely discussed underlying source responsible for both the emission of relativistic plasma and the production of FRBs is a young highly magnetized neutron star (magnetar) \citep[][]{2013Popov-FRB-Magnetar,Kulkarni+2014_FRB,Lyubarsky14maser,Katz16SGR,2016LuKumar-FRB-Magnetar,2016Murase-FRB-Magnetar,Beloborodov17Magnetar,Metzger17Magnetar,2017Nicholl-FRB-Magnetar} formed by a supernova explosion, embedded in expanding SN ejecta that confine the relativistic plasma \citep{2016Murase-FRB-Magnetar,Kashiyama17,Metzger17Magnetar}. Both magnetic and rotational energy are considered drivers of FRB and PRS emission in this scenario. Alternative scenarios for the formation of  systems with compact objects producing FRBs associated with PRSs include accretion-induced collapse \citep{Kashiyama17,2017Waxman-FRB,2018Metzger-BNS-AIC-FRB}, binary neutron star mergers \citep{2018Metzger-BNS-AIC-FRB}, run-away mass transfer in X-ray binaries \citep{2022Metzger-XRayBinary-FRB}, and white-dwarf mergers \citep{Kremer21,Kirsten22,2026Naoz-WD2-FRB}.

The temporal evolution of the dispersion measure (DM) predicted by most PRS models is inconsistent with that observed for FRB~121102, the most extensively monitored object, which shows a DM rise followed by a more rapid decline on a several-year timescale. The DM is predicted to evolve in most models due to the evolution of the shock waves arising from the interaction of the SN ejecta (or wind in the case of accreting binaries) with the surrounding medium, typically predicting a rapid DM decline over tens of years followed by a shallower DM decline or rise over hundreds to thousands of years \citep[e.g.][]{2018Piro-SN_DM,2022Metzger-XRayBinary-FRB,2026Murase-FRB-PRS-Magnetar}. An exception is the model discussed in \cite{2017Waxman-FRB} (hereafter Paper I), which predicted a rising DM due to the propagation of a shock driven by the relativistic PRS plasma into the confining cold plasma. We show here that the rapid decline observed at later times, and in particular the ratio of rise and decline rates, is consistent with the behavior expected once the shock wave crosses the confining plasma shell.

We note that \citet{2026Cui-RepeatingFRB-DMtrends} showed that, among the 19 repeating FRBs with more than ten bursts detected by CHIME \citep{2018CHIME-FRB-Overview,2021CHIME-FRB-Catalog}, $\approx26\%$ ($\approx11\%$) show a statistically significant secular decrease (increase) in DM. Secular DM evolution is therefore not a unique feature of FRB\,121102, and secular DM increase is not rare compared to DM decrease.

Before proceeding with the analysis, it should be mentioned that alternatives to the confined relativistic plasma model may be viable. The PRS emission may be produced by a relativistically expanding plasma, e.g., by a relativistic plasma wind driven by an active galactic nucleus black hole, with emission dominated by a compact region near the source where the relativistic electron density and magnetic field energy density are high. While observations disfavor the association of FRB~121102 with an AGN \citep{2023Hallinan-FRB-PRS-No-AGN}, this option cannot yet be ruled out.

The structure of this paper is as follows.
In \S\,\ref{sec:radio_constraints} we derive the key constraints on the source properties that are implied by the radio observations. The key implications to the nature of the source are summarized in \S\,\ref{sub-sec:implications}. In \S\,\ref{sec:SN} we discuss the DM evolution expected for a PRS confined by a supernova ejecta. We first discuss the case of homogeneous ejecta, showing that the DM is dominated by the shock driven into the ejecta by the PRS plasma (rather than by shocks arising from interactions with the ISM, which have been discussed previously) and that the resulting DM evolution is inconsistent with the observations; we then show that a highly clumpy ejecta may reproduce the observed behavior, provided that a large fraction of the ejecta mass resides in very small, dense clumps. In both sections, \S\,\ref{sec:radio_constraints} and \S\,\ref{sec:SN}, we also address the impact of ionizing radiation potentially emitted by the PRS. A detailed analysis of the DM measurements of FRB\,121102 is given in \S\,\ref{sec:DM-measurements}. In \S\,\ref{sec:scenarios} we discuss potential progenitor systems. Our results are summarized and discussed in \S~\ref{sec:discussion}.

Appendix~\ref{app:detailed} generalizes the DM evolution analysis of \S~\ref{sub-sec:DMvar} to a homologously expanding shell with a power-law density profile. In Appendices~\ref{app:numeric1} and~\ref{app:numeric2} we derive approximate analytic solutions that accurately describe the propagation of shock waves driven into a uniform density, homologously expanding ejecta by the injection of a relativistic plasma at the center at a constant energy injection rate, for the case where the onset of energy injection is delayed with respect to the onset of expansion (Appendix~\ref{app:numeric1}), and the case where the onset of energy injection coincides with the onset of expansion (Appendix~\ref{app:numeric2}).

\section{Source properties derived from radio observations}
\label{sec:radio_constraints}

We consider the setting of a relativistic magnetized plasma sphere confined by a colder denser shell with a power-law density dependence on radius and a homologous expansion velocity profile, $v\propto r$. The increase in DM is due to shock propagation in the shell, which ionizes the cold shell plasma, and the decrease is due to expansion following shock crossing of the initially cold shell. We assume that the confining shell is truncated at some finite radius, which is a valid approximation when the density declines rapidly beyond this radius, so that the mass beyond the truncation radius is small compared to that at smaller radii. In this case, the plasma beyond the truncation radius does not significantly affect the dynamics. Since the model is highly simplified, we do not follow a chi-square-based approach to fitting model parameters to the data (see further discussion in \S~\ref{sec:fit}).
Nevertheless, the simple model provides a good fit to the DM data, as shown in  Fig.~\ref{fig:DM_evolution} and discussed in \S~\ref{sec:fit}.

Our analysis and its conclusions hold for confining shells with a radial extent, $\Delta R$, that is not much smaller than the shell's radius, $R$. It does not apply to cases where the shell's DM, and hence mass, is dominated by a narrow $\Delta R \ll R$ sub-shell. Such a scenario is unlikely for two reasons. First, the mass ejected from an underlying source is expected to expand to $\Delta R\sim R$ unless it is ejected with a velocity that far exceeds the speed of sound at ejection. Second, the narrow shell scenario would require an unlikely fine-tuning/coincidence, as explained at the end of \S~\ref{sub-sec:implications}.

We ignore DM variations that may arise from changes in our line of sight to the source due to possible motion of the source in a binary system, which may be indicated by the observations suggesting a quasi-periodic modulation of the FRB activity with a period of $\sim160$\,d \citep{2020Rajwade-Periodicity,2021Cruces-Periodicity}.
The binary orbit required for variations over a few years is three orders of magnitude smaller than the minimum radius of the PRS bubble, $\approx10^{17}$~cm, set by the low level of scintillation variability \citep{2017Waxman-FRB,2023Hallinan-FRB-PRS-No-AGN}. Since the DM is likely produced by the non-relativistic plasma surrounding the PRS bubble, the binary motion is unlikely to lead to significant DM variations.

The constraints derived in Paper I are summarized in \S~\ref{sub-sec:PaperIconst}, the new constraints implied by the measurement of DM variations are derived in \S~\ref{sub-sec:DMvar}, the combined constraints are given in \S~\ref{sub-sec:PRS}, and the possible impact of photo-ionization is discussed in \S~\ref{sub-sec:photo-ion}. The implications for the FRB and PRS source are summarized in \S~\ref{sub-sec:implications}, where we also comment on the impacts of deviations from spherical symmetry and the possibility of instantaneous, rather than continuous, energy release from the underlying source driving the PRS.

\subsection{Constraints derived in Paper I}
\label{sub-sec:PaperIconst}

The observed properties used in Paper I to derive constraints on the PRS source are as follows.
\begin{enumerate}
  \item The luminosity and angular distances to the source, assuming that it resides in the dwarf galaxy, are $d_L=970$~Mpc and $d_A=680$~Mpc, respectively.
  \item The VLBI angular size, 0.2 and 2~mas at 5 and 1.7~GHz respectively \citep{Marcote17size}, is consistent with broadening due to scattering, $\theta\propto\nu^{-11/5}$. 
  \item At 3~GHz, the source shows 10\% to $30$\% variability on $\sim10$\,d time scale \citep{Chatterjee17FRB_loc} \citep[improved constraints are given in][]{2023Hallinan-FRB-PRS-No-AGN}.
  \item The radio flux peaks at $\sim10$\,GHz, with $\nu F_\nu\cong2\times10^{-17}$\,erg\,cm$^{-2}$\,s$^{-1}$, corresponding to $\nu L_\nu\cong2\times10^{39}$\,erg\,s$^{-1}$ \citep{Chatterjee17FRB_loc}.
  \item The flux extends approximately like $\nu F_\nu\propto \nu^1$ down to $\sim1$\,GHz \citep{Chatterjee17FRB_loc,Marcote17size}. This power-law extension was later observed to extend down to lower frequency, $0.4$\,GHz \citep{2021Vink-121102PRS-spec}.
  \item The total DM of the FRBs is $\approx 560$\,pc\,cm$^{-3}$ \citep{Spitler16Repeat,Chatterjee17FRB_loc}, and the contribution of the local-FRB environment is $\lesssim200$\,pc\,cm$^{-3}$ \citep{2017ApJ...834L...7T}. This is consistent with our analysis in \S~\ref{sec:fit} of the new DM data, which shows DM evolution- we find that the local contribution exceeds $70$\,pc\,cm$^{-3}$ with a best fit of $\approx200$\,pc\,cm$^{-3}$.
\end{enumerate}

The frequency-dependent angular size and limited flux variations that may be attributed to scintillation set an upper and a lower limit, respectively, on the PRS source radius, 
\begin{equation}
    \label{eq:R-limits}
    10^{17}\,{\rm cm}\lesssim R \lesssim 10^{18}\,{\rm cm.}
\end{equation}
The Lorentz factor of the relativistic electrons, $\gamma_{e}$, is constrained by the non-detection of self-absorption and cooling,
\begin{equation}\label{eq:g_e}
    \gamma_{e,2.5}>0.8t_{p,9}^{1/3},\quad \gamma_{e,2.5}>1.7 R_{17.5}^{-2},
\end{equation}
where $\gamma_{e}=10^{2.5}\gamma_{e,2.5}$, $t_p=10^9t_{p,9}$~s is the age of the persistent source, and $R=10^{17.5}R_{17.5}$~cm is the persistent source radius (note that the second limit, implied by the absence of self-absorption, is more stringent than obtained in Paper I, since the $\nu L_\nu\propto \nu^1$ behavior was observed to extend to lower frequency). 
The observed peak luminosity and peak frequency imply, using eqs. (10) and (11) of Paper I, that the energy $E_r$ carried by the relativistic electrons and the magnetic field of the PRS is
\begin{equation}\label{eq:E_r}
    E_{r,49}= 0.9\left( 0.4\gamma_{e,2.5}^3 + 0.6R_{17.5}^3\gamma_{e,2.5}^{-4} \right),
\end{equation}
where the first and second terms are the electron and magnetic field energy, respectively, and $E_r=10^{49}E_{r,49}$\,erg. 

Further constraints are inferred in Paper I from upper limits on DM variations. The detection of DM variations now allows improved constraints, as derived below.

\subsection{Constraints from DM variations}
\label{sub-sec:DMvar}

\begin{figure}
\centerline{\includegraphics[width=8cm]{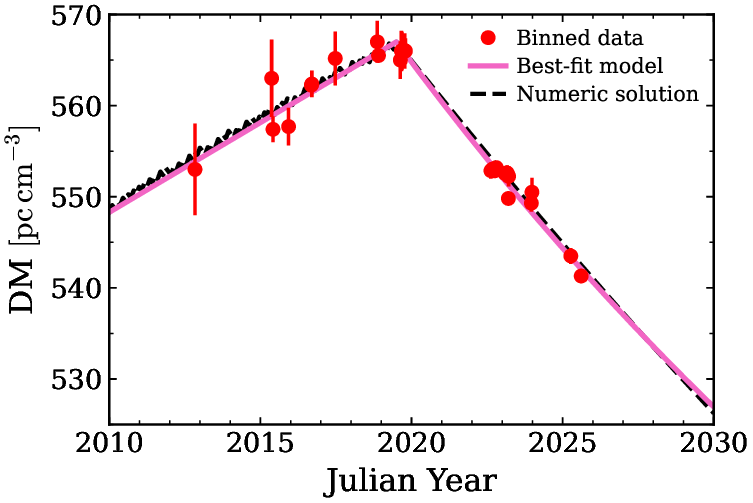}}
\caption{The DM evolution of FRB\,121102. Red points: the binned DM measurements of Table~\ref{tab:FRB20121102_BinnedDM}, with the errors described in \S~\ref{sec:Data}. Dashed black line: The DM obtained from a numeric calculation of a shock driven into a homologously expanding plasma by the injection of relativistic plasma at the center with a constant energy injection rate, for $v_0/v_s=0.09$ ($\eta=10^4$, see Eq.~(\ref{eq:eta})) and dimensional parameter values given following Eq.~(\ref{eq:v_b_an}). Red line: the best-fit analytic model, Eq.~(\ref{eq:DM_Model}), with external DM $d_{\rm ext}\approx360\,{\rm cm^{-3}\,pc}$, peak total DM $d_{\rm peak}\cong567\,{\rm cm^{-3}\,pc}$, $t_{\rm s}=R/v_s\approx2.8\times10^{9}$\,s, $t_0=R/v_0\approx3.7\times10^{10}$\,s, $t_{\rm peak}\cong2019.5$\,JY and $\alpha\approx1.8$.
The DM rises approximately linearly with time while the PRS-driven shock propagates through the confining shell, and declines approximately as the inverse of the shell's radius squared after the shock crosses it.
\label{fig:DM_evolution}}
\end{figure}

A detailed analysis of the DM variations is given in \S~\ref{sec:DM-measurements}. The DM rises before the peak by approximately $\Delta DM_-=13$\,pc\,cm$^{-3}$ in $\Delta t_-=7$\,yrs, and decreases by approximately $\Delta DM_+=-25$\,pc\,cm$^{-3}$ in the following $\Delta t_+=6$\,yrs. In this section, we use only the observed DM rise and decline rates. Nevertheless, as noted above, the DM evolution predicted by the simple model provides a good fit to the full DM data (see \S~\ref{sec:fit} and Fig.~\ref{fig:DM_evolution}).

We first present simple approximate analytic estimates, and then demonstrate their validity and accuracy by comparison with exact (1D) hydrodynamic calculations. In the analytic analysis, we assume that the distances traveled by the shock and then by the expanding shell during the observation time are small compared to the shell's radius, and we verify the consistency of this assumption with the obtained results.

Consider the shock wave driven by the high pressure of the relativistic plasma into the surrounding cold plasma. Let us first assume, for simplicity, that the cold plasma is of uniform density, that its expansion speed is negligible (relative to the shock speed), and that the shock velocity is time-independent. As the shock propagates through the cold plasma, it ionizes and compresses it. The total number of post-shock free electrons is $N_e=(4\pi/3) r_s^3\chi_e n_i$, where $r_s$ is the shock radius, $n_i$ is the pre-shock ion density, and $\chi_e$ is the number of free electrons per ion. The shocked plasma is compressed behind the shock to a narrow shell, $dr/r=(1/3)(\rho_1/\rho_2)\ll1$ where $\rho_2/\rho_1\ge4$ is the shock compression factor (neglecting the ionization energy, $\rho_2/\rho_1=4$). Thus, at times preceding the breakout of the shock through the outer plasma edge at radius $R$, the time dependence of the DM is approximately given by
\begin{equation}
    \label{eq:DMt-}
    DM_-=\frac{N_e}{4\pi r_s^2}=\frac{1}{3}\chi_e n_i r_s,\quad
    \dot{DM}_-=\frac{1}{3}\chi_e n_i v_s,
\end{equation}
where $v_s$ is the shock velocity. After the shock breaks through the cold plasma, the DM decreases due to the post-shock expansion, 
\begin{equation}
    \label{eq:DMt+}
    DM=\frac{1}{3}\chi_e n_i \frac{R}{(1+vt/R)^2},\quad
    \dot{DM}_+=-\frac{2}{3}\chi_e n_i v,
\end{equation}
where $R$ is the initial outer shell radius, $v$ is the post-shock expansion velocity, and the breakout time is chosen as $t=0$. Estimating the expansion velocity through energy conservation, assuming that the post-shock internal energy is converted to kinetic energy, we have $v\approx 2\sqrt{2}v_s/(\gamma+1)\approx v_s$ where $\gamma=5/3$ is the adiabatic index of the plasma, and $\dot{DM}_+/\dot{DM}_-=-2$. The shell's expansion may be accelerated by the internal pressure. However, the change in velocity over the observation time will be small under our assumption of small shell expansion.

Under the assumption that the distances traveled by the shock and the expanding shell are small compared to $R$, we may approximate $\dot{DM}_+/\dot{DM}_-=(\Delta DM_+/\Delta t_+)/(\Delta DM_-/\Delta t_-)=-2.2$. The fact that the measured value is very close to, but slightly larger in magnitude than, $-2$ expected for the case of a breakout through a static shell, implies both that this model is consistent with the observations and that the expansion velocity of the pre-shocked plasma, $v_0$, should be much smaller than $v_s$. This is due to the fact that a significant expansion speed both reduces $\dot{DM}_-$ and increases $|\dot{DM}_+|$. 

To obtain an analytic constraint on $v_0/v_s$, and also examine the effect of non-uniform density, we generalize in Appendix~\ref{app:detailed} the above analysis to the case of a homologously expanding plasma with a power-law density profile, $\rho\propto r^{-\alpha_\rho}$. We show that for this case
\begin{equation}\label{eq:DMrat}
  X_{DM}\equiv-\frac{\dot{DM}_+}{\dot{DM}_-}=2\frac{(1+v_0/v_s)}{(1-\alpha_\rho-2v_0/v_s)},
\end{equation}
implying that
\begin{equation}\label{eq:v_rat}
    \frac{v_0}{v_s}=\frac{X_{DM}(1-\alpha_\rho)-2}{2(1+X_{DM})}< 0.04.
\end{equation}
The conclusion that $v_0/v_s\ll1$ is not sensitive to the exact value of $X_{DM}$. Increasing the value of $X_{DM}$ from 2.2 to 2.8, for example, changes the upper limit on $v_0/v_s$ from 0.04 to 0.1.

Denoting by $\Sigma$ the maximal DM produced by the shocked plasma, $\Sigma=\chi_e n_i R/3$, we have
\begin{equation}
    \label{eq:ni}
    \chi_e n_i=2.9\times10^3 \Sigma_2 R_{17.5}^{-1},
\end{equation}
where $\Sigma=100\Sigma_2$\,pc\,cm$^{-3}$. Recall that the data imply $\Sigma_2>1$ with a best fit of $\Sigma_2=2$. From the measured values of $\dot{DM}$ we infer
\begin{equation}\label{eq:v_S}
  v_s=\frac{\dot{DM}_-}{\Sigma}
  R= 1.9\times10^8 \Sigma_2^{-1}R_{17.5} \,{\rm cm\,s}^{-1}.
\end{equation}
The inferred velocity implies that our assumption of small shock and shell propagation distances over the observation time, $\Delta t=4\times10^8$\,s, is valid, $v_s\Delta t/R=0.2\Sigma_2^{-1}$.

The post-shock pressure, $p_s=(3/4)n_i A m_p v_s^2$ where $A$ is the atomic mass number, is 
\begin{equation}\label{eq:p}
  p_s=1.3\times10^{-4}(A/\chi_e) \Sigma_2^{-1} R_{17.5}\,{\rm erg\,cm}^{-3},
\end{equation}
and the energy of the hot relativistic plasma, $4\pi p_s R^3$, is
\begin{equation}\label{eq:E_rs}
  E_r=5.2\times10^{49} (A/\chi_e) \Sigma_2^{-1} R_{17.5}^4\,{\rm erg}.
\end{equation}
The age of the persistent source is $\sim R/v_s$. For the case of a time-independent rate of energy injection (in the form of relativistic plasma) at the center, $t_p = (3/5)R/v_s$ (see Appendix~\ref{app:numeric1}),
\begin{equation}\label{eq:t_p}
  t_p= 1.0\times 10^9 \Sigma_2\,{\rm s},
\end{equation}
implying a rate of energy injection to the relativistic plasma of
\begin{equation}
    \label{eq:Edot}
    \dot{E}_r=5.2\times10^{40} (A/\chi_e) \Sigma_2^{-2} R_{17.5}^4 {\rm erg/s}.
\end{equation}
The mass of the initially cold plasma, $4\pi R^2\Sigma (A/\chi_e) m_p$, is
\begin{equation}\label{eq:M}
  M = 0.32 (A/\chi_e) \Sigma_2  R_{17.5}^2\, {\rm M}_\odot.
\end{equation}

For the inferred shock velocity, $\sim 10^8$\,cm\,s$^{-1}$, the post-shock temperature is high, $\sim10$~keV, compared to the ionization energies (of H-O), and the shocked plasma is expected to be highly ionized, implying $A/\chi_e$ in the range of 1-2. The ionization energy is also much smaller than the incoming ion kinetic energy, hence it may be neglected in calculating the shock Hugoniot relations.

In Appendix~\ref{app:numeric1} we derive an approximate analytic solution that accurately reproduces the results of 1D hydrodynamic calculations of the propagation of a shock wave into a cold plasma sphere with uniform density expanding homologously, driven by the injection of a highly relativistic plasma at the sphere's center with a constant energy injection rate, $\dot{E}$. The flow is completely defined by a single dimensionless parameter,
\begin{equation}
    \label{eq:eta}
    \eta\equiv\frac{\dot{E}R_0/v_0}{E_k}=\frac{5}{2\pi}\frac{\dot{E}}{\rho_0R_0^2v_0^3},
\end{equation}
where $R_0$ is the initial radius of the sphere, $v=(r/R_0)v_0$ is the initial velocity field, $\rho_0$ is the initial density, and $E_k=0.3(4\pi/3)\rho_0 R_0^3 v_0^2$ is the initial kinetic energy. Note that the dependence on the dimensional parameters, $R_0$, $v_0$ and $\rho_0$ is straightforwardly obtained by dimensional considerations. For large $\eta$, the shock radius is given by
\begin{equation}
\label{eq:rs_an}
r_s =  0.80 \left(\frac{\dot{E}}{\rho_0}t^3\right)^{1/5},
\end{equation}
and the shock velocity at breakout is given by
\begin{equation}
    \label{eq:v_b_an}
    \frac{v_{s,b}}{v_0}=0.45\eta^{1/3}.
\end{equation}

Figure~\ref{fig:DM_evolution} shows the time-dependent DM obtained from the numeric solution for $\eta=10^4$ corresponding to $v_0/v_s=0.09$. The dimensional parameters required to obtain $\Sigma_2=2$ and $13$\,pc\,cm$^{-3}$ DM increase during 7\,yrs before the peak are $M = 0.55\,{\rm M}_\odot$ and $v_0=1.1\times10^7$\,cm\,s$^{-1}$ (for $R_0=10^{17.5}$\,cm), implying $v_s=1.3\times10^8$\,cm\,s$^{-1}$,  $\dot{E}=1.7\times10^{40}$\,erg\,s$^{-1}$ and relativistic plasma energy at breakout of $E_{r}=2.1\times10^{49}$\,erg. The model DM is consistent with observations, and the analytic parameter values, $v_s$, $M$ and $E_r$, derived above are in good agreement with those obtained numerically. Both the numeric and analytic values of $v_0/v_s$ are $\ll1$, as required in order to obtain $\dot{DM}_+/\dot{DM}_-\approx-2$. The numeric value, $0.09$, is, however, larger than the $0.04$ value obtained by the simplified analytic analysis that relies on a simplified description of the hydrodynamics.

\subsection{Combining the constraints}
\label{sub-sec:PRS}

Eqs.~(\ref{eq:E_rs}) and~(\ref{eq:E_r}) imply
\begin{equation}\label{eq:Req}
    6(A/\chi_e)\Sigma_2^{-1} R_{17.5}^4 = 0.4\gamma_{e,2.5}^3 + 0.6R_{17.5}^3\gamma_{e,2.5}^{-4}.
\end{equation}
The two branches of the solution, for electron and magnetic field energy domination, respectively, are
\begin{eqnarray}
    \label{eq:gR}
    \gamma_{e,2.5}&=&2.5(A/\chi_e)^{1/3}\Sigma_2^{-1/3}R_{17.5}^{4/3}, \nonumber\\ 
    \gamma_{e,2.5}&=&0.56(A/\chi_e)^{-1/4}\Sigma_2^{1/4}R_{17.5}^{-1/4},
\end{eqnarray}
and
\begin{equation}\label{eq:Rmin}
    R\ge1.6(\Sigma_2\chi_e/A)^{7/19}\times 10^{17}\,{\rm cm},
\end{equation}
where the minimum value of $R$ is obtained for $\gamma_{e,2.5}=2^{1/7}R_{17.5}^{3/7}$ and corresponds to e--B equipartition ($U_B/U_e=0.75$). The first requirement of Eq.~(\ref{eq:g_e}) implies that for larger $R$ the solution lies on the electron energy domination branch, with $U_B/U_e\propto R^{-19/3}$ (and $E_e\propto R^4$). Using the second requirement of Eq.~(\ref{eq:g_e}) implies $R\gtrsim2.7\times10^{17}(\Sigma_2\chi_e/A)^{1/10}$\,cm and hence $E_r\gtrsim3(\Sigma_2\chi_e/A)^{-0.6}\times10^{49}$\,erg.

\subsection{Photo-ionization}
\label{sub-sec:photo-ion}

The energy distribution of the relativistic electrons, which are responsible for the PRS emission and are likely "heated" by collisionless shocks, may include a non-thermal component extending to high energies, leading to the emission of ionizing radiation. Assuming that a fraction $f_{\rm NT}\sim0.1$ of the electron energy is carried by a power-law distribution of electron energies, $\gamma_e^2 dn_e/d\gamma_e={\rm const.}$, the ionizing radiation luminosity is
\begin{equation}
    \label{eq:L_NT}
    \nu L_{\nu,\rm NT}=0.5\frac{f_{\rm NT}}{\ln(\nu_{\max}/\nu_{\min})}\dot{E}_r\approx 3\times 10^{37} f_{\rm NT,-1} \dot{E}_{r,40} {\rm erg\,s}^{-1}
\end{equation}
where $f_{\rm NT}=10^{-1}f_{\rm NT,-1}$, $\dot{E}_r=10^{40}\dot{E}_{r,40}$\,erg\,s$^{-1}$ and we have assumed $\nu_{\max}/\nu_{\min}=10^8$ (The fraction of relativistic plasma energy carried by electrons is large in the allowed range of parameters, see \S~\ref{sub-sec:PRS}). We note that the observed PRS spectrum between 10 and 30\,GHz declines approximately as $\nu L_\nu\propto \nu^{-0.2}$ \citep{2021Vink-121102PRS-spec}. If the spectrum follows this power-law behavior to $\sim10^{16}$\,Hz it would yield an ionizing flux of $\approx10^{38}$\,erg\,s$^{-1}$.

Let us examine whether the ionization of the confining plasma may be dominated by ionizing radiation rather than by the relativistic PRS plasma-driven shock. For a highly ionized Hydrogen dominated plasma, the case-B radiative electron recombination rate (the inverse electron life time) is given by $t_{\rm RR}^{-1}=n_i <\sigma v_e>_{\rm RR}$ with $<\sigma v_e>_{\rm RR}=2.7\times10^{-13}T_{\rm eV}^{-3/4}{\rm cm^3/s}$ where $T=1T_{\rm eV}$\,eV is the electron temperature \citep[e.g.,][]{2011Draine-book} and $n_i=3\Sigma/\chi_e/R$ is the ion density,
\begin{equation}
    \label{eq:RRrate}
    t_{\rm RR}^{-1} = 8\times10^{-10} \Sigma_2 \chi_e^{-1} R_{17.5}^{-1}T_{\rm eV}^{-3/4} {\rm s^{-1}}.
\end{equation}
The ionizing luminosity required to maintain equilibrium ionization, $4\pi R^2 \Sigma t_{\rm RR}^{-1} I$ where $I$ is the ionization energy, is 
\begin{equation}
    \label{eq:L_RR}
    L_{\rm RR}
    \approx 2\times10^{36} \Sigma^2_2 R_{17.5}\,{\rm erg\,s}^{-1},
\end{equation}
approximating the equilibrium temperature by $T_e\approx I/3$. Note that if the recombination time is long compared to the source age, $t_p$, the minimum luminosity required for full ionization is $4\pi R^2 \Sigma t_{p}^{-1} I$. In the case under consideration, $t_p\sim t_{\rm RR}$, hence the two requirements would yield a similar limit on the required luminosity. Comparing eqs.~(\ref{eq:L_RR}) and~(\ref{eq:L_NT}), we find that if the confining plasma is H dominated, the ionizing radiation that may be produced by the PRS may lead to complete ionization of the confining plasma prior to its ionization by the propagating shock wave, thus suppressing the time variation of the DM due to shock propagation.

Let us consider next a heavier plasma composition. For an Oxygen rich plasma with significant ionization, $\chi_e=3,4$, the ionization energies are $\approx100$\,eV, and for $T_e\approx I/3\approx30$\,eV recombination is dominated by dielectronic recombination with $<\sigma v_e>_{\rm DR}\approx10^{-11}$\,cm$^3$\,s$^{-1}$ \citep{2023CHIANTI,2003Badnell-Recomb}.
This implies
\begin{equation}
    \label{eq:DRrate}
    t_{\rm DR}^{-1} \approx 10^{-8} (\chi_e/3)^{-1}\Sigma_2 R_{17.5}^{-1}\,{\rm s^{-1}},
\end{equation}
and the ionizing luminosity required for maintaining such a high ionization level is
\begin{equation}
    \label{eq:L_DR}
    L_{\rm DR}
    \approx 6\times10^{38} (\chi_e/3)^{-1}\Sigma^2_2 R_{17.5}\,{\rm erg\,s}^{-1}.
\end{equation}
Comparing eqs.~(\ref{eq:L_DR}) and~(\ref{eq:L_NT}), we find that for a heavy composition, the ionization that may be produced by the possible ionizing radiation of the PRS is likely sub-dominant to that produced by the propagating shock wave. 

\subsection{Implications}
\label{sub-sec:implications}

\begin{enumerate}
    \item \textit{Source energetics.} The energy of the PRS plasma is $\gtrsim10^{49.5}$\,erg and its age is $\approx2\times10^9$\,s, implying an underlying source power $\gtrsim10^{40}$\,erg\,s$^{-1}$.
    \\ If the underlying energy source is a neutron star, this implies a rotational rather than a magnetic energy source. The neutron star magnetic field required to carry the total energy exceeds $10^{16}$\,G, and while such a field could be dissipated over hundreds of years by ambipolar diffusion, this dissipation occurs in high-density regions, where the dissipated energy is carried by neutrinos and escapes \citep{1992Goldreich-AmbipolarDiff,2016Beloborodov-MagnetarHeating}. The more likely energy source is rotational. The inferred energy requires a period of $\approx10$\,ms ($E_{\rm rot}\approx2\times10^{50}(P/10\,{\rm ms})^{-2}$\,erg), and energy extraction over $\sim100$\,yr requires $B\sim10^{12.5}$\,G (the dipole spin-down time is $\approx10^{3}(B/10^{12}{\rm G})^{-2}(P/10\,{\rm ms})^{2}$\,yr) \citep[e.g.,][]{1983Shapiro-Teukolsky-book}. An alternative energy source may be accretion ($\dot{M}\approx2\times10^{-6}(\dot{E}_r/10^{40}$\,erg\,s$^{-1}$)($\epsilon/0.1)^{-1}\,{\rm M}_\odot$/yr  where $\epsilon$ is the efficiency). 
    \item \textit{An age hierarchy.} The expansion velocity of the confining plasma is constrained to be much smaller than the shock velocity, $v_0/v_s\lesssim0.1$. This implies, in particular, that the onset of the expansion of the confining plasma occurred well before the onset of the emission of the relativistic plasma at the center. That is, the age of the PRS is $t_p=30\Sigma_2$\,yr, and the ejecta expansion time, $t_0=R/v_0$ exceeds $10 R/v_s=10(5/3)t_p=500\Sigma_2$\,yr. 
    \item \textit{No energetic stellar explosion.} The implied expansion velocity of the confining plasma, $v_0\le 200 (10v_0/v_s) \Sigma_2^{-1} R_{17.5}\,{\, \rm km\,s^{-1}}$, is well below the typical $5000$\,km\,s$^{-1}$ speed of SN ejecta. The mass of the confining plasma is $M=0.32(A/\chi_e)\Sigma_2R_{17.5}^2M_\odot$ (eq.~\ref{eq:M}), i.e. $\approx0.3$--$3\,{\rm M}_\odot$ over the allowed range of $R$, and its kinetic energy, $\approx 10^{47} (A/\chi_e)\Sigma_2^{-1} (10v_0/v_s)^2 R_{17.5}^4$\,erg, is well below the $10^{51}$\,erg typical for SN ejecta. The ejection of the confining plasma was therefore not due to an energetic stellar explosion. 
    \item \textit{Shell radius and density profile.} The confining shell's outer radius lies in the range of $3-10\times10^{17}$\,cm and its density profile must be significantly shallower than $\rho\propto r^{-1}$.
    \item \textit{Shell composition.} If the confining plasma is H-dominated and the PRS is efficient in producing ionizing radiation, photo-ionization may dominate ionization by the propagating shock wave, thus suppressing the time variation of the DM due to shock propagation. For a heavier, e.g., C/O dominated, composition, photo-ionization is likely sub-dominant. The observation of a large DM variation therefore disfavors a hydrogen-dominated confining plasma, unless the PRS is an inefficient producer of ionizing radiation.
\end{enumerate}

In the analysis above, we have assumed spherical symmetry and a wide initial radial extent of the confining shell, $\Delta R\sim R$. For large-scale deviations from spherical symmetry or large-scale density inhomogeneities that do not introduce small dimensionless parameters in the characterization of the ejecta, the results derived above will hold qualitatively, with order-unity modifications to the numerical values. Below, we briefly consider the possible impacts of confining shell density distributions that introduce small parameters.

Before discussing the introduction of small parameters, let us comment on the validity of our assumption that the post-shock expansion flow is radial (spherical expansion) in the presence of large-scale deviations of the density distribution from a spherical one. The post-shock expansion may deviate from a radial flow field in case that the density distribution is not spherical and $v_s\gg v_0$, "overriding" the radial homologous expansion. A cylindrical structure of high density, shocked by the inner pressure to $v_s\gg v_0$, may, for example, lead to a cylindrical post-shock expansion. In such a scenario, the numerical values of inferred parameters will be modified, while the conclusion $v_0/v_s\ll1$ remains valid.

As noted at the opening of \S~\ref{sec:radio_constraints}, we have not considered a scenario in which the DM is dominated by a narrow shell of thickness $d\ll R$, or by a narrow ``patch'' of thickness $d\ll R$ and lateral extent $L\gg d$. We note here that such scenarios require an unlikely coincidence to account for the observations. The DM rise time is the shock crossing time of the narrow shell/patch\footnote{If the density contrast between the narrow shell and the interior is sufficiently large, $\gg(R/d)^2$, the shock crossing time $d/v_s$ may exceed the homologous expansion time, $R/v_0$. In this case, the DM increase that arises as the shock enters the shell will be followed, after a time $\sim R/v_0$, by a decrease due to expansion before the shock crosses the shell. The transition from DM rise to decline will be smooth, with $\dot{DM}=0$ at the peak DM time, and the (absolute) decline rate will be lower than the rise rate. This contrasts with the observed sharp transition to a decline at a rate greater than the rise rate.}, $d/v_s$, while the DM decline time is set by the lateral expansion of the shocked plasma: $R/v_0$ if the post-shock expansion is dominated by the homologous expansion ($v_s\ll v_0$), and $L/v_s$ (patch) or $R/v_s$ (shell) if it is dominated by the shock-induced expansion ($v_s\gg v_0$). In the latter case, the rise time is much shorter than the decline time, since $d\ll L,R$, in contradiction with the observed similar rise and decline times. In the former case, similar rise and decline times require a coincidence of two unrelated small numbers, $d/R\approx v_s/v_0$. This should be contrasted with the extended-shell model considered in this paper, where both time scales are $\sim R/v_s$ and their similarity is natural.

The case of a clumpy ejecta, with high-density regions characterized by $L\sim d\ll R$, is discussed in detail in \S~\ref{sub-sec:SN clumpy}. We show that the constraints 2 and 3 may be evaded in this case if the DM variations are dominated by a shock compression of a single dense clump that happens to lie along our line of sight. For the SN ejecta case, however, the required clumping properties make this scenario unlikely.

Finally, a note is in place regarding the energy injection. We have assumed continuous, time-independent energy injection in the form of a relativistic plasma at the center. An impulsive energy release, of $\sim10^{49.5}$\,erg in a mildly relativistic plasma as considered in Paper I, would lead to similar shock propagation and DM evolution as obtained for the continuous energy release case. However, such an instantaneous release would produce the relativistic PRS plasma only if the region surrounding the source is "evacuated" up to a radius $\sim R$, so that it contains a mass much smaller than that ejected by the instantaneous event. Otherwise, the released energy would be transferred to the shock wave propagating through the cold, non-relativistic confining plasma, leaving little energy in the relativistic plasma.

\section{SN ejecta DM}
\label{sec:SN}

In this section, we show that in the scenario where the PRS is confined by a SN ejecta with typical parameters, the observed DM behavior may be produced if much of the ejecta mass is carried by a very large number, $\sim 10^5$, of over-dense and very small, $10^{-2.5}$ fractional size, clumps. We open with a discussion of the DM produced by a homogeneous spherical ejecta, since the dominant contribution to the DM, which is produced by the shock driven by the PRS plasma into the ejecta, was largely neglected in earlier analyses.

\subsection{Homogeneous ejecta}
\label{sub-sec:SNhomog}

In Appendix~\ref{app:numeric2} we derive an approximate analytic solution that accurately describes the propagation of a shock wave driven into a uniform density, homologously expanding SN ejecta by the injection of a relativistic plasma at the center at a constant energy injection rate, assuming that the onset of energy injection at the center is coincident with the SN explosion time. We show that the shock radius, $r_s$, is given by
\begin{equation}
    \label{eq:rsSN}
    \frac{r_s}{R}=\xi\left(\frac{\dot{E}t}{E_k}\right)^{1/5},
\end{equation}
where $R=v_0t$ is the ejecta outer radius, $\dot{E}$ is the energy injection rate, $E_k$ is the initial ejecta kinetic energy, and $\xi$ is a constant close to unity, given in the analytic solution to be $\xi=0.85$. The DM produced by the shock-ionized part of the ejecta is
\begin{eqnarray}
    \label{eq:SN-DM}
    DM&=&\frac{r_s}{R}\frac{(\chi_e/A)M/m_p}{4\pi R^2}
    =2.3\times10^5\frac{\xi\chi_e}{A}\left(\frac{M}{10M_\odot}\right)^{4/5}\nonumber\\ &\times&\left(\frac{v_0}{5000{\, \rm km\,s^{-1}}}\right)^{-12/5}\dot{E}^{1/5}_{40}t_{\rm yr}^{-9/5}\,{\rm pc\,cm}^{-3}. 
\end{eqnarray}

The shock driven by the PRS into the SN ejecta thus leads to a rapidly declining DM, $\dot{DM}=-(9/5)DM/t\approx-{\rm few}\times10^{2}$\,pc\,cm$^{-3}$\,yr$^{-1}$ at an age of $t=10$\,yr. In order to allow an increase in DM, $\dot{DM}\approx1$\,pc\,cm$^{-3}$\,yr$^{-1}$ as observed, due to another process (e.g., clumpiness as discussed in the following sub-section), the age should be sufficiently large so that the absolute decline rate implied by eq.~(\ref{eq:SN-DM}) drops below $\approx1$\,pc\,cm$^{-3}$\,yr$^{-1}$, i.e. $t>100$\,yr or, equivalently, $R>10^{18}\,$cm.

Let us next consider the possible impact of PRS ionizing radiation. For a hydrogen-dominated ejecta, the case-B electron radiative recombination rate is
\begin{equation}
    \label{eq:SN-RRrate}
    t_{\rm RR}^{-1} = 0.7\times10^{-4}\frac{M}{10M_\odot}\left(\frac{v_0}{5000{\, \rm km\,s^{-1}}}\right)^{-3}t_{\rm yr}^{-3} {\rm s^{-1}},
\end{equation}
implying $t_{\rm RR}<t$ for $t\lesssim50$~yr. The radius $R_I$ out to which the ionizing luminosity maintains high ionization is given by 
$\nu L_\nu=(4\pi/3) R_I^3 (\rho/m_p) t_{\rm RR}^{-1} I$, implying
\begin{eqnarray}
    \label{eq:SN-DM-I}
    DM_I&=&\frac{1}{3}\frac{\rho}{m_p}R_I=
    2.4\times10^4 \left(\frac{M}{10M_\odot}\frac{\nu L_{\nu,NT}}{3\times10^{37}{\rm erg/s}}\right)^{1/3}\nonumber\\ &\times&
    \left(\frac{v_0}{5000{\, \rm km\,s^{-1}}}\right)^{-1}t_{\rm yr}^{-1}\,{\rm pc\,cm}^{-3}.    
\end{eqnarray}
Comparing eqs.~(\ref{eq:SN-DM}) and~(\ref{eq:SN-DM-I}), we find that photoionization may dominate the ionization of the inner part of the ejecta at $t>20$\,yr (For heavier composition ejecta, the shock-driven ionization would dominate to longer times). The conclusion that $t>100$\,yr, $R>10^{18}\,$cm, is required for the absolute decline rate of the DM to drop below $\approx 1$\,pc\,cm$^{-3}$\,yr$^{-1}$, remains valid in this case too. 

The DM produced by the ionization of the inner part of the ejecta is larger than the commonly discussed contribution to the DM by the ionization of the outer part of the ejecta and of the surrounding ISM, due to the shocks driven into the ISM and back into the ejecta,
\begin{eqnarray}
    \label{eq:SN-DM-ISM}
    DM_{\rm ISM}&\approx& 10^2 \left(\frac{M}{10M_\odot}\frac{n_{\rm ISM}}{1{\rm cm}^{-3}}\right)^{1/2}\nonumber\\ &\times&
    \left(\frac{v_0}{5000{\, \rm km\,s^{-1}}}\right)^{-1/2}t_{\rm yr}^{-1/2}\,{\rm pc\,cm}^{-3}.
\end{eqnarray}

\subsection{Highly clumpy ejecta}
\label{sub-sec:SN clumpy}

At an age of 100 years, which is the minimal age required to avoid too rapid DM decrease for $v_0=5000\,{\rm km\,s}^{-1}$, the ejecta expands to a radius $R=1.6\times10^{18}$\,cm, and the PRS-driven shock reaches (see eq.~(\ref{eq:rsSN})) $r_s=6.3\times10^{17}$\,cm for $\dot{E}_r=10^{40}$\,erg\,s$^{-1}$ and $M=10$\,M$_\odot$. 

The observed increase, followed by a decrease in the DM, is inconsistent with the evolution predicted for homogeneous ejecta. It may be accounted for, however, by assuming highly clumpy ejecta. Consider a small clump of size $d\ll R$ with a large over-density $\rho'/\rho\gg1$. Due to its high density, the PRS-driven shock propagates rapidly around it, producing high pressure at the clump boundary, which drives a converging shock into the clump. If our line of sight passes through a single clump, we will observe a rising DM as the shock compresses and ionizes the clump plasma, followed by a decreasing DM as the outer-most location of the shock on the plane of the sky crosses our line of sight. From this time onward, the converging flow removes ionized plasma from our line of sight (at late times, when the shock "bounces" and expands, another rise in DM is expected). 

Let us estimate the values of $d$ and $\rho'/\rho$ required to produce a $\Delta DM=20$\,pc\,cm$^{-3}$ rise and drop in DM over $\Delta t=6$\,yrs periods. Estimating $\Delta DM=\rho' d/m_p$, and $\Delta t=d/v_s'$ with $v_s'=\sqrt{p/\rho'}$, we have
\begin{equation}
    \label{eq:d-clump}
    d\approx\frac{p(\Delta t)^2}{m_p\Delta DM}=3\times10^{15}\,{\rm cm},
\end{equation}
and 
\begin{equation}
    \label{eq:rho-clump}
    \rho'\approx 3\times10^{-20}\,{\rm g\,cm}^{-3},
    \quad \frac{\rho'}{\rho}\approx 30,
\end{equation}
where we have used $p=E_r/(4\pi r_s^3)$ with $E_r=3\times10^{49}$\,erg and $r_s=6.3\times10^{17}$\,cm. 

In order for the clump interpretation to be self-consistent, the shock velocity, $v_s'=d/\Delta t$, must be much larger than the clump's expansion velocity, $(d/R)v_0=d/t$, i.e. $\Delta t/t\ll1$ is required. This is consistent with the scenario discussed above, where $\Delta t/t=0.06$, and implies that a $\sim10$\% coincidence between our observing time and the clump-shocking time is required, beyond the requirement that the line-of-sight should cross a clump. 

The fraction of the ejecta mass that must be enclosed in the clumps in order for them to cover a significant fraction of the surface area of the ejecta, hence providing a significant probability for a random line of sight to cross a clump, is $4(d/r_s)(\rho'/\rho)\approx 0.5$. Thus, in order for the clump explanation to be probable, a significant fraction of the ejecta mass must reside in dense clumps of very small, $\approx10^{-2.5}$, fractional size, making this scenario unlikely.

\section{DM measurements}
\label{sec:DM-measurements}

\subsection{The DM data}
\label{sec:Data}

We collected a total of 2285 dispersion measurements (DM) from the literature (\citealt{2014Spitler-121102DM, Spitler16Repeat, 2019Hessels-121102DM, 2023Snelders-121102DM, 2020Oostrum-121102DM, 2021Platts-121102DM, 2020Majid-121102DM, 2021Li-121102DM, 2025Wang-121102DM}).
Since this work heavily relies on this data set, we list all these measurements in Table~\ref{tab:FRB20121102_AllDM}.
\begin{deluxetable}{lll}
\tablecolumns{3}
\tablewidth{0pt}
\tablecaption{All DM measurements of FRB\,121102}
\tablehead{
\colhead{MJD}    &
\colhead{DM}   &
\colhead{$\Delta$DM} \\
\colhead{(day)}       &
\colhead{(cm$^{-3}$\,pc)}       &
\colhead{(cm$^{-3}$\,pc)}       
}
\startdata
56233.2828370080  & 553.000 &  5.00\\
57159.7376008350  & 560.000 &  2.00\\
57159.7442236190  & 566.000 &  5.00\\
57175.6931432320  & 555.000 &  2.00\\
57175.6997278260  & 558.000 &  6.00
\enddata
\tablecomments{All DM measurements of FRB\,121102. Collected from: \cite{2014Spitler-121102DM, Spitler16Repeat, 2019Hessels-121102DM, 2023Snelders-121102DM, 2020Oostrum-121102DM, 2021Platts-121102DM, 2020Majid-121102DM, 2021Li-121102DM, 2025Wang-121102DM}. The full table is available from the electronic version of the paper. Here, the first five lines are listed.}
\label{tab:FRB20121102_AllDM}
\end{deluxetable}
 
Next, we binned the DM in 10 days bins, where the bins were
measured relative to the first data point (Julian year 2012.83504). In each bin, we took the median of all measurements as the representative point and the error was obtained from the standard deviation (StD) of the measurements in each bin. We did~not divide the errors by the square root of the number of points in each bin. The motivation here is that the scatter may represent some additional intrinsic or extrinsic noise that cannot be ignored in the statistical analysis. Typically, this StD is in the range of $0.3$\,cm$^{-3}$\,pc to $2$\,cm$^{-3}$\,pc. In cases in which only one data point is in the bin, we used its reported error. In case that a bin contains a single DM measurement, we also added to the DM error, in quadrature, an error of 0.5\,cm$^{-3}$\,pc. This was needed in order to make the $\chi^{2}$ per degree of freedom of our best fit model, about unity (see details below). Most authors converted the times of bursts to 0 dispersion and TDB barycentric time\footnote{We assumed this is the case also for papers, where this information is not disclosed.}. However, different authors measure the DM using different methods, and hence, this can lead to small variations in DM and time of arrival. This may introduce yet another uncertainty, both in the DM and its time of arrival. Table~\ref{tab:FRB20121102_BinnedDM} provides the binned DM along with their estimated errors.

\begin{deluxetable}{llll}
\tablecolumns{4}
\tablewidth{0pt}
\tablecaption{Binned DM data for FRB\,121102}
\tablehead{
\colhead{Julian Year}    &
\colhead{DM}   &
\colhead{$\Delta$DM} &
\colhead{$N_{\rm pt}$} \\
\colhead{(JY)}       &
\colhead{(cm$^{-3}$\,pc)}       &
\colhead{(cm$^{-3}$\,pc)}       &
\colhead{()}    
}
\startdata
      $2012.83582$  &  $553.00$  &  $ 5.02$  &  $  1$  \\ 
      $2015.37232$  &  $563.00$  &  $ 4.24$  &  $  2$  \\ 
      $2015.41610$  &  $557.40$  &  $ 1.41$  &  $  7$  \\ 
      $2015.93211$  &  $557.70$  &  $ 2.06$  &  $  1$  \\ 
      $2016.68478$  &  $562.24$  &  $ 0.73$  &  $  3$  \\ 
      $2016.70273$  &  $562.38$  &  $ 1.45$  &  $ 12$  \\ 
      $2017.47611$  &  $565.18$  &  $ 2.95$  &  $  2$  \\ 
      $2018.86754$  &  $567.00$  &  $ 2.29$  &  $ 27$  \\ 
      $2018.90743$  &  $565.50$  &  $ 0.71$  &  $  2$  \\ 
      $2019.62835$  &  $565.00$  &  $ 2.06$  &  $  1$  \\ 
      $2019.66354$  &  $566.00$  &  $ 2.20$  &  $510$  \\ 
      $2019.68494$  &  $566.00$  &  $ 2.10$  &  $340$  \\ 
      $2019.72301$  &  $566.10$  &  $ 1.46$  &  $310$  \\ 
      $2019.74350$  &  $566.10$  &  $ 1.60$  &  $446$  \\ 
      $2019.77133$  &  $566.00$  &  $ 1.93$  &  $ 23$  \\ 
      $2019.79845$  &  $566.00$  &  $ 1.38$  &  $ 29$  \\ 
      $2022.62607$  &  $552.86$  &  $ 0.87$  &  $  7$  \\ 
      $2022.67699$  &  $553.03$  &  $ 0.55$  &  $230$  \\ 
      $2022.70839$  &  $553.02$  &  $ 0.48$  &  $127$  \\ 
      $2022.72964$  &  $553.04$  &  $ 0.42$  &  $ 58$  \\ 
      $2022.74852$  &  $552.84$  &  $ 0.65$  &  $  1$  \\ 
      $2022.80304$  &  $553.20$  &  $ 0.37$  &  $  5$  \\ 
      $2023.09533$  &  $552.55$  &  $ 0.53$  &  $ 20$  \\ 
      $2023.13604$  &  $552.61$  &  $ 0.67$  &  $ 30$  \\ 
      $2023.16744$  &  $552.59$  &  $ 0.86$  &  $ 48$  \\ 
      $2023.20258$  &  $549.81$  &  $ 0.55$  &  $  1$  \\ 
      $2023.21298$  &  $552.24$  &  $ 1.11$  &  $ 33$  \\ 
      $2023.96439$  &  $549.30$  &  $ 0.94$  &  $  3$  \\ 
      $2023.98353$  &  $550.50$  &  $ 1.58$  &  $  1$  \\ 
      $2025.26947$  &  $543.50$  &  $ 0.86$  &  $  3$  \\ 
      $2025.61349$  &  $541.30$  &  $ 0.71$  &  $  2$  
\enddata
\tablecomments{The binned DM data of FRB\,121102. 
$N_{\rm pt}$ is the number of points in the bin.
$\Delta$DM is the error in the DM calculated from the std in the bin (without dividing by $\sqrt{N_{\rm pt}}$. Furthermore, for bins with $N_{\rm pt}=1$, we added in quadrature 0.5\,cm$^{-3}$\,pc to $\Delta$DM.}
\label{tab:FRB20121102_BinnedDM}
\end{deluxetable}

\subsection{Model fitting}
\label{sec:fit}

Motivated by the simple model derived from the physical constraints on the
PRS size and radio emission (see Appendix \S~\ref{app:detailed}), we fitted a simple model of the form
 \begin{equation}
\mathrm{DM}(t)= d_{\rm ext}+d_{\rm src}
\begin{cases}
\left[1+t\left(\dfrac{1}{t_{\rm s}}-\dfrac{2}{t_0}\right)\right],
& t \le 0, \\[1.2em]
\left[1+\left(\dfrac{1}{t_{\rm s}}+\dfrac{1}{t_0}\right)t\right]^{-\alpha},
& t>0,
\end{cases}
\label{eq:DM_Model}
 \end{equation}
Here 
$t=t_{\rm JY} - t_{\rm peak}$, $t_{\rm JY}$ is the time of the measurements in Julian years, $t_{\rm peak}$ is the time of peak DM, $d_{\rm ext}$ is the external DM (i.e., not from the FRB or PRS), $d_{\rm src}=d_{\rm peak}-d_{\rm ext}$ where $d_{\rm peak}$ is the value of DM at peak, $t_{s}=R/v_{s}$ and $t_0=R/v_0$, and $\alpha$ is a parameter that captures the impact of the geometry of the post-shock expansion- $\alpha=2$ for spherical expansion and $\alpha=1$ for cylindrical expansion (with line of sight perpendicular to the cylinder's axis). 

As explained in \S~\ref{sec:radio_constraints}, the post-shock expansion may deviate from a radial flow (spherical expansion) only for $v_s\gg v_0$ ("overriding" the radial homologous expansion, \S~\ref{sub-sec:implications}). This implies that $\alpha<2$ is acceptable only for $t_s\ll t_0$. In addition, the phenomenological form of the time dependence chosen in Eq.~(\ref{eq:DM_Model}) is expected to hold at short times, over which the distances traveled by the shock and expanding shell are small (\S~\ref{sub-sec:DMvar}), i.e., for $|t|\ll t_s$ (this is required to justify the linear $t$-dependence approximation at $t<0$ and for neglecting acceleration by the internal pressure at $t>0$). This implies that the results obtained are valid and self-consistent for $t_s$ values much larger than the observation time before and after the peak in DM, $t_s\gg2\times10^8\,$s. In our fitting below of the parameters of the model of Eq.~(\ref{eq:DM_Model}) to the data, we do not impose the requirements $t_s\gg2\times10^8\,$s and $t_s\ll t_0$ for $\alpha<2$, in order to allow the exploration of a wider range of parameter values. We find that these constraints are nevertheless satisfied by the resulting preferred parameter values.

We fitted the six free parameters of the problem:
(i) the external DM $d_{\rm ext}$;
(ii) the peak DM $d_{\rm peak}$;
(iii) the time of peak DM ($T_{\rm peak}$);
(iv) the time scale of the shock going through the cold gas ($t_{\rm s}$);
(v) the expansion time scale of the cold gas ($t_0$);
and (vi) the geometry parameter $\alpha$.
We scan the entire parameter space
with $d_{\rm ext}$ in the range of 300\,cm$^{-3}$\,pc to 550\,cm$^{-3}$\,pc, with steps of 10\,cm$^{-3}$\,pc;
$d_{\rm peak}$ in the range of 565\,cm$^{-3}$\,pc to 572\,cm$^{-3}$\,pc with steps of 0.5\,cm$^{-3}$\,pc;
$T_{\rm peak}$ in the range of 2018 to 2020.2, with steps of 0.1;
$t_{\rm s}$ in 30 logarithmic steps between $10^{8.5}$\,s to $10^{11}$\,s; $t_0$ in 70 logarithmic steps between $10^{8.5}$\,s to $10^{16}$\,s; and $\alpha$ in the range of 1 to 2 with 0.1 steps. Time is measured in Julian years relative to the J2000.0 epoch.

Since our physical model is highly simplified (see \S~\ref{sec:radio_constraints}), 
we do~not quote the best fit parameters with their formal uncertainty. 
Instead, we provide rough acceptable regions defined by $\Delta\chi^{2}\approx9$.

The best fit $\chi^{2}$/dof is $55/25$, which is not too bad given the simple nature of this model. In order to fix this $\chi^{2}$/dof, we added to the errors of the binned DM measurements that contain only a single measurement in a bin, $0.5$\,cm$^{-3}$\,pc, in quadrature. Here, the motivation is that the standard deviation in the DM measurements in each bin may better represent the uncertainties.

The global minimum parameter values are:
$d_{\rm ext}\approx360$\,cm$^{-3}$\,pc, $d_{\rm peak}\cong567$\,cm$^{-3}$\,pc, $t_{\rm s}\approx2.8\times10^9$\,s, $t_0\approx3.7\times10^{10}$\,s, 
$t_{\rm peak}\cong2019.5$\,JY, and $\alpha\approx1.8$.
In Figure~\ref{fig:DM_evolution} we present the binned DM measurements as well as the best-fit model. The parameter values obtained from the fit are in excellent agreement with those obtained in the analytic analysis of \S~\ref{sec:radio_constraints} (note that $t_p=3t_s/5$). Performing a similar parameter fit with fixed $\alpha=2$ yields essentially the same results.

\begin{figure}
\centerline{\includegraphics[width=8cm]{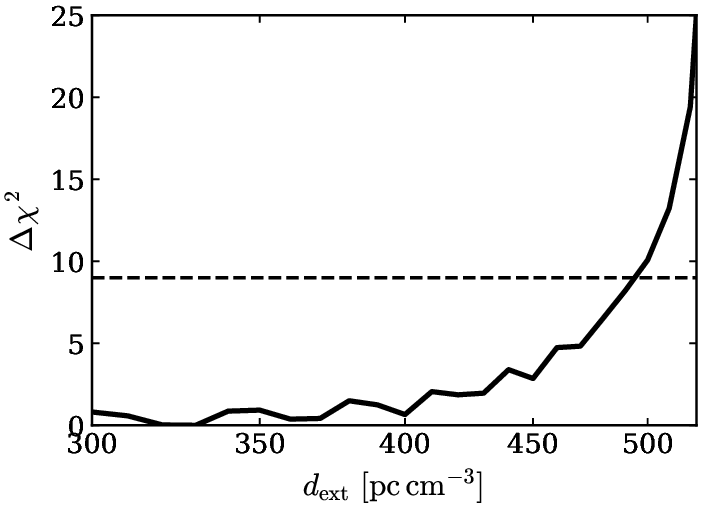}}
\caption{$\Delta\chi^2$ as a function of the model parameter (eq.~(\ref{eq:DM_Model})) $d_{\rm ext}$, marginalized over all the other parameters. The horizontal line marks $\Delta\chi^2=9$, adopted as a rough estimate of the acceptable parameter range (\S~\ref{sec:fit}).
\label{fig:DM_FitPar}}
\end{figure}

\begin{figure}
\centerline{\includegraphics[width=8cm]{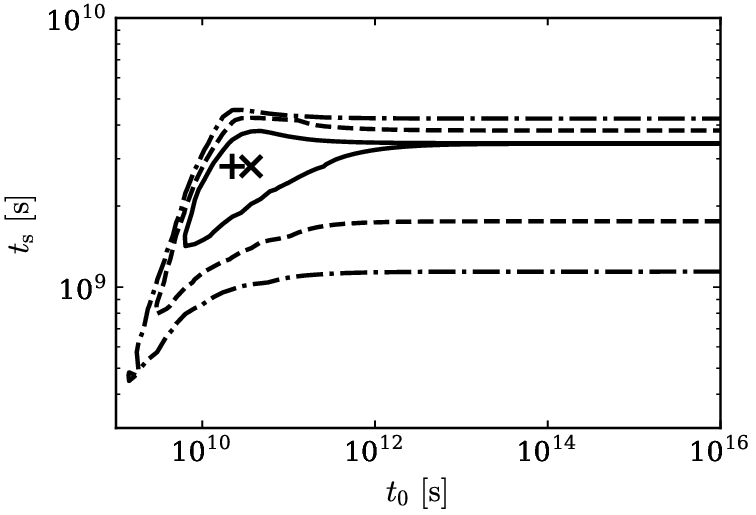}}
\caption{Formal $1$, $2$, $3$-$\sigma$ error regions solid, dashed, dashed-dotted lines, respectively; for two degrees of freedom) in the $t_{\rm s}$ vs. $t_0$ plane, where all the other parameters were marginalized. $\times$ marks the global best fit position, while $+$ marks the minimum $\Delta\chi^{2}$ in this specific marginalized plot.
\label{fig:DM_ts_tcold}}
\end{figure}

In Figure~\ref{fig:DM_FitPar} we present the $\Delta\chi^{2}$ of $d_{\rm ext}$ 
when marginalized over all the other parameters.
Adopting the $\Delta\chi^{2}\approx 9$, as a rough estimate for the valid range of the parameters, we find (from the marginalized parameters)
$d_{\rm ext}<490$\,cm$^{-3}$\,pc. We also show, in Figure~\ref{fig:DM_ts_tcold}, the $\Delta\chi^{2}$ as a function of $t_{\rm s}$ and $t_0$ marginalized over all the other parameters. The best fit values give $v_s=1.1\times10^3R_{17.5}\,{\, \rm km\,s^{-1}}$, $v_0=85R_{17.5}\,{\, \rm km\,s^{-1}}$, $v_0/v_s=t_s/t_0=0.08$, and $\Delta\chi^{2}$ rises sharply for $v_0/v_s=t_s/t_0>0.25$. The fit allows, within $\Delta\chi^{2}\approx 9$, $t_s=5\times10^8$~s and  $t_0=2\times10^9$~s corresponding to $v_0\approx5\times10^3(R/10^{18}\,{\rm cm})\,{\, \rm km\,s^{-1}}$. 
Limiting the parameter range to $t_s>10^9$ (recalling that self-consistency requires $t_s\gg2\times10^8\,$s), $v_0\approx2.5\times10^3(R/10^{18}\,{\rm cm})\,{\, \rm km\,s^{-1}}$ is allowed within $\Delta\chi^{2}\approx 9$, implying that a typical SN ejecta velocity is excluded at $>3\sigma$ even for the largest allowed $R$. 


\section{Possible scenarios for the progenitor system}
\label{sec:scenarios}

In this section, we consider possible astrophysical scenarios for the
formation of the system, in light of the constraints on source properties derived in \S~\ref{sec:radio_constraints} and summarized in
\S~\ref{sub-sec:implications}.
We examine progenitors classes that we find most relevant -- the collapse of a massive star (\S~\ref{sec:massive}), a double white dwarf (WD) merger followed by a delayed collapse to a neutron star (NS) (\S~\ref{sec:WDWD}), and a common envelope ejection followed by a delayed compact-object--core merger (\S~\ref{sec:CE}) -- and comment briefly on other possibilities (\S~\ref{sec:other}). Since detailed model calculations of the relevant mass-ejection processes involve many uncertain ingredients, we deliberately rely below only on robust, order-of-magnitude characteristics of each scenario and refer to published detailed estimates as indicative only. The scenarios are compared
in \S~\ref{sec:scenario_summary}, where we argue that the WD merger scenario is the most promising one.

We use the constraints on source properties given in \S~\ref{sub-sec:implications}. We note that the constraint (5) on shell composition may be evaded if the PRS is an inefficient producer of ionizing radiation, but it discriminates sharply between the scenarios considered. As explained in \S~\ref{sec:SN}, the key constraints may be avoided for a highly clumpy SN ejecta. However, the required clumping properties, with most of the ejecta mass carried by $\sim 10^5$ clumps of very small, $\sim 10^ {-2.5}$, fractional size, makes this scenario less likely in our view.

\subsection{Collapse of a massive star}
\label{sec:massive}

In this class of scenarios, the compact object is formed in the core collapse of a massive star. A regular SN explosion is excluded since the confining shell velocity and energy are well below those typical of SN ejecta. 
The slow shell must therefore be provided in this case either by the pre-collapse wind of the star, or by the ejection of a shell of very low energy and mass accompanying the collapse, which must produce a BH rather than a NS to avoid the formation of a typical SN ejecta. A possible reservation regarding this scenario is that the collapse forms a BH, and while accretion can comfortably supply the PRS energetics, attributing the repeating coherent FRBs to a BH may be less natural than to a NS. In addition, the ejected envelope is hydrogen-rich, which may be in tension with the composition constraint (5).

{\it (a) Pre-collapse wind.} A slow massive wind, e.g. $\dot{M}_w=3\times10^{-5}\,M_\odot$\,yr$^{-1}$ with $v_w=10$\,km\,s$^{-1}$ lasting $10^{4}$\,yr -- within the range observed for red supergiants and for eruptive mass loss of evolved massive stars (see \citealt{Smith2014ARA&A_TypeIIn_SN_MassLoss} for a review) -- places $M_w=\dot{M}_w R/v_w\approx0.3\,{\rm M}_\odot$ within $R=3\times10^{17}$\,cm, consistent with the requirements of \S~\ref{sub-sec:implications}. The required time delay between mass ejection and the creation of the PRS (requirement 2 of \S~\ref{sub-sec:implications}), is natural here: the wind is emitted during the final $\sim10^4$\,yr of the star's life, and the engine -- accretion onto the newly formed BH -- turns on at collapse. The wind scenario is challenged by the requirement that the confining shell density should decrease with radius much more slowly than $\rho\propto r^{-1}$. This does not directly exclude the wind scenario, as the mass-loss rate may be variable and episodic.

{\it (b) A failed SN with a low-energy envelope ejection.} If the star
collapses to a BH without an explosion, the near-instantaneous
reduction of the core gravitational mass by neutrino emission
($\sim0.2-0.5\,M_\odot$ over a few seconds) launches a weak shock
that may unbind the loosely bound envelope of a red supergiant
\citep{Nadyozhin80,Lovegrove+Woosley2013ApJ_FailedSupernovae_Theory_LightCurves}. The simulations of
\citet{Lovegrove+Woosley2013ApJ_FailedSupernovae_Theory_LightCurves} find ejection of $0.1-0.5\,{\rm M}_\odot$ with
$E_k\approx10^{46}-6\times10^{47}$\,erg and characteristic velocities
of $\sim100-300$~km/s (see also \citealt{Piro13}); later calculations
find similar energies with ejecta masses reaching several
$M_\odot$ (at correspondingly lower velocities) for red supergiant
progenitors \citep{Fernandez+2018_FailedSN_massejection}, which would match the upper end of
the allowed range of $R$ and $M$. A candidate for such an event --
a $\sim25\,M_\odot$ red supergiant that disappeared following a weak
optical transient -- may have already been observed
\citep{Adams+Kochanek+2017_FailedSN_Candidate}. While the detailed numbers should be regarded
as uncertain, the gross features -- ejection of a few tenths of a
solar mass at a few hundred km/s, in a roughly spherical shell -- is
a direct consequence of the (robust) neutrino mass loss and of the
weak binding of the envelope, and matches requirements of \S~\ref{sub-sec:implications} remarkably well. The match is best for the lower end of the allowed radius range, $R\approx2\times10^{17}$~cm ($M\approx0.13(A/\chi_e)\Sigma_2\,M_\odot$, $E_k\approx2\times10^{47}$~erg), in which the shell reaches $R$ roughly $500$~yr after the collapse.

The principal difficulty of this scenario is that the envelope ejection and the BH formation are {\it simultaneous}, so accretion power is available immediately, whereas the observations require the engine (and the PRS) to be activated only $\approx{\rm few}\times10^{2}$~yr later. A delay of this magnitude does not emerge naturally from the collapse physics and must be inserted in an ad hoc manner (e.g., by postulating that fallback accretion becomes super-critical, or that a jet manages to escape, only centuries after the collapse). 

\subsection{A double WD merger with a delayed collapse}
\label{sec:WDWD}

In this scenario the system is formed by the merger of two WDs whose total mass exceeds the Chandrasekhar mass, with the merger remnant collapsing to a NS
a long time after the merger. Which WD mergers end in such a collapse -- rather than in a thermonuclear explosion or a stable massive WD -- is not reliably known; it has been suggested that super-Chandrasekhar Carbon-Oxygen (CO)+CO mergers with sufficiently unequal masses (for which carbon ignites off-center and burns the remnant to Oxygen-Neon (ONe) rather than exploding it), as well as mergers involving ONe WDs, follow this path (e.g. \citealt{Yoon07,Shen12,Schwab16,Kremer26}, and references therein). We do not rely on these specific assignments; for our purpose, it suffices that some WD mergers plausibly produce an NS long after the merger. The scenario then naturally supplies all the required components.

{\it The slow shell.} Mass can be lost from the system at low velocity at several stages of the merger: dynamically, during the tidal disruption of the secondary; during the subsequent viscous evolution of the remnant; and via winds blown from the hot, extended, near-Eddington remnant during its thermal (Kelvin-Helmholtz) phase, which may last $\sim10^{3}-10^{4}$~yr \citep{Shen12,Schwab16}. The dynamically ejected tidal-tail material is estimated at only $\sim10^{-3}-10^{-2}\,{\rm M}_\odot$, moving at the WD escape velocity, $\sim10^{3}$\,km\,s$^{-1}$ \citep{RaskinKasen13} -- both too little and too fast to constitute the required confining shell -- so the shell must be dominated by the slower outflows of the viscous and thermal phases. None of these mass-loss channels has been reliably calculated, and we regard the published estimates with due caution; the robust point is the velocity scale, set by the escape velocity from the outer edge of the
inflated remnant ($\sim10^{13}$\,cm), which is $\sim{\rm few}\times10$\,km\,s$^{-1}$ \citep{Shen12} -- comfortably below the $v_0$ limit. Material leaving the remnant at such velocities over the thermal phase reaches $10^{17}-10^{18}$\,cm by the time of collapse -- the required radius. A crude upper bound on the wind mass, from
momentum conservation for a radiation-driven outflow, $\dot{M}_{\rm wind}\sim L_{\rm Edd}/(c\,v_{\rm esc})\sim10^{-5}\,{\rm M}_\odot$\,yr$^{-1}$ over $\sim10^{4}$\,yr, gives
$\sim0.1\,{\rm M}_\odot$; together with the dynamically ejected mass, this is consistent with constraints on the shell's radius and mass. 
Additional mass loss during the final, rapid mass-transfer phase preceding the merger has also been suggested (e.g. \citealt{Inoue26}), though such calculations again involve many uncertain ingredients. 

The radial structure of the outflow, and hence how the requirement for a shallow density profile (decreasing slower than $\rho\propto r^{-1}$) is met, depends on which mass-loss channel dominates. While mass ejected promptly, during the viscous phase, may be characterized by a shallow profile, 
a quasi-steady wind blowing throughout the thermal phase up to the collapse may produce a wind-like, $\rho\propto r^{-2}$ profile. This does not directly exclude this scenario, since, as in the massive-star wind scenario, the mass loss may be both variable and episodic. 
Separately from the radial structure, merger outflows are expected to be concentrated toward the orbital plane; such angular asphericity is neither required nor constrained by the observations, as large-scale inhomogeneities are not expected to modify the constraints qualitatively (see \S~\ref{sub-sec:implications}). 

{\it The delay.} The remnant does not collapse at the merger. It must
first evolve through its viscous phase and then its thermal phase,
during which (in the picture of \citealt{Shen12,Schwab16}) an off-center
carbon burning front propagates to the center and converts the
remnant to ONe, with electron captures triggering the collapse only
thereafter; the estimated total delay is of order
$10^{3}-10^{4}$\,yr. While the quantitative time scale is uncertain
(it depends on poorly constrained viscous and burning physics), the
{\it existence} of a long delay between the mass ejection (at and
shortly after the merger) and the formation of the NS is a generic,
qualitative feature of the scenario, with a magnitude compatible
with $t_0\approx1.5\times10^{3}(R/3\times10^{17}{\rm
cm})(v_0/60\,{\, \rm km\,s^{-1}})^{-1}$\,yr. 

{\it The engine and the hot plasma.} The AIC of a rapidly rotating
merger remnant is expected to form a rapidly rotating
($P\lesssim10$\,ms) NS with a modest magnetic field, for which the
rotational energy, $E_{\rm rot}\approx2\times10^{50}(P/10\,{\rm
ms})^{-2}$\,erg, and the dipole spin-down time, $\sim10^{3}(B/10^{12}{\rm
G})^{-2}(P/10\,{\rm ms})^{2}$\,yr, match the required $E_r$ and
$\dot{E}_r$; accretion of leftover disk/envelope
material, $\sim10^{-3}\,{\rm M}_\odot$ over the PRS lifetime, is an
alternative with similar energetics. 
The young, rapidly rotating NS may serve as the repeating FRB source. Young NSs formed in an AIC (of an accreting WD or of a WD merger remnant) have indeed been proposed as the source of FRB~121102 and of repeating FRBs generally \citep{Kashiyama17,Margalit19}, on grounds independent of the DM-variation constraints derived here. We note that the association of the repeating FRB~20200120E with a globular cluster in M81 -- an old stellar environment where core collapse does not occur -- has independently motivated WD-merger/AIC formation channels for repeating FRB sources \citep{Kirsten22,Kremer21}.

Finally, we note that an impulsive energy release of $\sim10^{49.5}$\,erg in a mildly relativistic ejecta may accompany the last stage of collapse to a NS \citep[e.g.,][]{Dessart+2006_AIC,Metzger09,Kashiyama17}. As noted in \S~\ref{sub-sec:implications}, this would lead to similar shock propagation and DM evolution as obtained for the continuous energy release of $\sim10^{49.5}$\,erg over $\sim10^9$~s. The relativistic PRS plasma would need to be produced by a subsequent continuous energy emission from the underlying source, since impulsively released energy would be almost entirely transferred to the shock wave propagating through the cold, non-relativistic confining plasma.

\subsection{Common envelope ejection with a delayed compact-object--core merger}
\label{sec:CE}

A third scenario that should be considered is a common envelope (CE) episode in a binary consisting of a giant star and a compact companion (a NS, a BH, or a WD). The CE phase generically ends in one of two ways: the envelope is ejected and a compact binary survives, or the envelope ejection fails and the companion spirals into the giant's core. Both the ejection and the in-spiral are poorly understood quantitatively, but two gross features are robust and are exactly what is needed here. First, the envelope, of mass $\sim0.1-1\,{\rm M}_\odot$ (or more), is released at a characteristic velocity of order the envelope escape velocity, $\sim10-10^{2}$\,km\,s$^{-1}$ \citep{Ivanova13}
(The geometry is expected to be equatorially concentrated; as in \S~\ref{sec:WDWD}, such angular asphericity is neither required nor constrained by the observations). This expectation is supported observationally: luminous red novae, which are believed to be CE ejections/stellar mergers caught in the act (the case of V1309~Sco being direct \citep{Tylenda11}), show ejecta of
$\sim10^{-2}-1\,{\rm M}_\odot$ at $\sim10^{2}$\,km\,s$^{-1}$ \citep{MacLeod17,MetzgerPejcha17}. Second, if the envelope ejection is incomplete, the orbit of the compact companion continues to decay -- through drag in the residual envelope material and/or tidal interaction -- until the companion merges with the core. The merger disrupts the
core and establishes an accretion flow onto the compact object, providing the delayed engine: the delay is the post-CE orbital decay time, which is set by the residual envelope mass and structure and can plausibly span the required $\sim10^{2}-10^{4}$\,yr. Mergers of a NS or a BH with the companion's core have indeed been considered as engines -- in the context of energetic, promptly interacting SNe \citep{Chevalier12,Soker19}; what we invoke here is a milder and delayed version of the same configuration.

We emphasize that this scenario is, at present, the least developed of the three considered. The key quantitative ingredient -- the post-CE orbital decay time of the compact companion through the residual envelope following a {\it failed} ejection -- has, to our knowledge, not been reliably calculated, and the $\sim10^{2}-10^{4}$\,yr range quoted above should be regarded as plausible rather than established. Likewise, while compact-object--core mergers have been considered engines of energetic, promptly interacting SNe, the delayed, low-power variant invoked here has not been studied in the literature, and no calculations exist for the resulting mass ejection, engine power, or longevity. The scenario should therefore be viewed as a qualitative possibility whose consistency with the constraints of \S~\ref{sub-sec:implications} cannot yet be robustly assessed.

The accretion of core material onto a NS companion at
$\dot{M}\sim10^{-6}-10^{-5}\,{\rm M}_\odot$\,yr$^{-1}$ comfortably supplies
the required energy release rate, and if the companion is a NS -- possibly spun up
and rejuvenated by the accretion -- the scenario also provides a
plausible repeating FRB source. The main open questions are
whether the delayed-merger channel (as opposed to prompt merger
during the CE, or successful ejection with no merger) is realized
with a reasonable probability, and whether the engine can remain as
steady as required over decades. 
The composition of the shell in this scenario is that of the giant's
envelope -- hydrogen-rich for a red giant donor, but hydrogen-poor
(He or C/O-rich) if the giant is itself stripped -- so the
composition constraint (5) of \S~\ref{sub-sec:implications} discriminates
between variants of this scenario as well.

\subsection{Other scenarios}
\label{sec:other}

For completeness we mention other channels that could in principle
eject $\sim0.1\,M_\odot$, and why we disfavor them.

{\it Compact object mergers.} NS-NS and NS-WD mergers eject too
little mass at too high a velocity ($\sim10^{-2}\,{\rm M}_\odot$ at
$\gtrsim0.1c$, and $\lesssim0.1\,{\rm M}_\odot$ at $\sim10^{4}$\,km\,s$^{-1}$,
respectively).

{\it Tidal disruption by an intermediate-mass BH.} The unbound debris
of a disrupted star is ejected at too large a velocity, $v\sim10^{4}$\,km\,s$^{-1}$; the bound debris is accreted on time scales too short compared to $t_0$.

\subsection{The most promising scenario}
\label{sec:scenario_summary}

Table~\ref{tab:scenarios} summarizes the confrontation of the scenarios with the requirements of \S~\ref{sub-sec:implications}. The massive-star scenarios reproduce the shell mass, velocity, and energy, 
but the failed SN provides no natural delay between mass ejection and engine activation. Both variants leave a BH rather than an NS as the putative burst source, and both predict a hydrogen-rich shell. The CE scenario shares the attractive delayed-engine structure and can accommodate an NS, but relies on a poorly constrained post-CE in-spiral. The double WD merger with a delayed collapse satisfies all the requirements without fine tuning: the merger and the subsequent evolution of the remnant plausibly shed $\sim0.1\,{\rm M}_\odot$ of C/O-rich material at $\sim10-10^{2}$\,km\,s$^{-1}$, which by the time of the delayed ($\sim10^{3}-10^{4}$\,yr) collapse occupies $R\sim10^{17}-10^{18}$\,cm; the collapse forms a rapidly rotating NS whose rotational (or accretion) energy matches the PRS energetics and which provides a young NS as the repeating FRB source. We therefore regard the WD+WD merger scenario as the most promising. Its distinctive predictions -- a hydrogen-free, metal-rich shell, an old-population host environment being permissible, and PRS/FRB activity decaying on the NS spin-down or fallback time scale of decades -- offer several avenues for testing it.

\begin{deluxetable*}{lcccccc}
\tablecolumns{7}
\tablewidth{0pt}
\tablecaption{Confronting the scenarios with the requirements of
\S~\ref{sub-sec:implications}\label{tab:scenarios}}
\tablehead{
\colhead{Scenario} & \colhead{(1)} & \colhead{(2)} & \colhead{(3)} &
\colhead{(4)} & \colhead{(5)} & \colhead{FRB source}
}
\startdata
Massive star: wind $+$ BH        & $+$ & $+$     & $+$     & $\circ$ & $-$     & $?$     \\
Massive star: failed SN          & $+$ & $-$     & $+$     & $+$     & $-$     & $?$     \\
WD$+$WD merger $+$ delayed collapse & $+$ & $+$  & $+$     & $\circ$ & $+$     & $+$     \\
CE ejection $+$ delayed merger   & $\circ$ & $\circ$ & $+$ & $+$     & $\circ$ & $\circ$ \\
NS-NS / NS-WD merger             & \nodata & \nodata & $-$ & \nodata & \nodata & \nodata \\
IMBH tidal disruption            & \nodata & \nodata & $-$ & \nodata & \nodata & \nodata
\enddata
\tablecomments{Columns (1)--(5) follow the numbering of \S~\ref{sub-sec:implications}: (1) source energetics; (2) an age hierarchy (delayed engine activation); (3) no energetic stellar explosion (slow massive shell); (4) shell radius and shallow density profile; (5) hydrogen-poor shell composition. The last column denotes the availability of a natural repeating-FRB source (a young, rapidly rotating and/or strongly magnetized NS; an accreting BH cannot be excluded). $+$: naturally satisfied; $\circ$: possible with additional
assumptions; $?$: uncertain; $-$: disfavored. The NS-NS/WD merger and IMBH tidal disruption models are ruled out due to the high velocity of the ejected plasma and are mentioned for completeness. }
\end{deluxetable*}

\section{Discussion}
\label{sec:discussion}

The secular DM evolution of FRB~20121102 provides a direct dynamical probe of the plasma confining its persistent radio source. The principal results are as follows.
\begin{itemize}
    \item \textit{The DM evolution.} The rise and subsequent decline of the DM are naturally explained by a shock driven by the relativistic PRS plasma through a cold confining shell, see Fig.~\ref{fig:DM_evolution}. Before breakout, the ionized electron column increases as the shock sweeps up additional material; after breakout, the column decreases approximately as the inverse square of the expansion radius. The observed ratio of the decline and rise rates is close to the value expected for a shell in which the initial expansion velocity is small compared with the shock velocity. This conclusion is insensitive to order-unity departures from spherical symmetry and to moderate radial density gradients.
    \item \textit{The PRS properties.}  Combining the DM and synchrotron constraints implies a compact, $(3$--$10)\times10^{17}$\,cm, energetic, $E_r\gtrsim10^{49.5}$\,erg, source (\S~\ref{sub-sec:PRS}) with a $t_p\approx2\times10^9$\,s age (Eq.~(\ref{eq:t_p})).
    \item \textit{The PRS energy source.} The PRS energetics favor rotation or accretion over magnetic-field decay as the underlying energy source. A magnetic reservoir exceeding $10^{49.5}$\,erg requires an internal field of order $10^{16}$\,G. Although the microphysics of field evolution is uncertain, dissipation at the densities that contain most of this energy is expected to be neutrino-dominated \citep{1992Goldreich-AmbipolarDiff,2016Beloborodov-MagnetarHeating}, and the efficient transfer of energy to a decades-long external relativistic nebula is challenging. A neutron star with $P\sim$ a few--10~ms and $B\sim10^{12}$\,G accounts for the energy and timescale.
    \item \textit{An age hierarchy.} The expansion velocity of the confining plasma is constrained to be much smaller than the shock velocity, $v_0/v_s\lesssim0.1$ (\S~\ref{sub-sec:DMvar}, \S~\ref{sec:fit}). This implies, in particular, that the onset of the expansion of the confining plasma occurred well before the onset of the emission of the relativistic plasma at the center. That is, the age of the PRS is $t_p=30\Sigma_2$\,yr, and the ejecta expansion time, $t_0=R/v_0$ exceeds $10 R/v_s=10(5/3)t_p=500\Sigma_2$\,yr (Recall that $\Sigma_2=2$ is the best fit value of the conifing shell column density). 
    \item \textit{No energetic stellar explosion.} The implied expansion velocity of the confining plasma, $v_0\le 200 (10v_0/v_s) \Sigma_2^{-1} R_{17.5}\,{\, \rm km\,s^{-1}}$ (\S~\ref{sub-sec:DMvar}), is well below the typical $5000$\,km\,s$^{-1}$ speed of SN ejecta. The mass of the confining plasma is $M=0.32(A/\chi_e)\Sigma_2R_{17.5}^2M_\odot$ (eq.~\ref{eq:M}), i.e. $\approx0.3$--$3\,{\rm M}_\odot$ over the allowed range of $R$, and its kinetic energy, $\approx 10^{47} (A/\chi_e)\Sigma_2^{-1} (10v_0/v_s)^2 R_{17.5}^4$\,erg, is well below the $10^{51}$\,erg typical for SN ejecta. The ejection of the confining plasma was therefore not due to an energetic stellar explosion. 
    \item \textit{SN DM evolution.} A PRS-powered shock propagating through freely expanding SN ejecta produces DM\,$\propto t^{-9/5}$ (for uniform ejecta and constant injection power), with a large negative $\dot{\rm DM}\approx-{\rm few}\times10^{2}$\,pc\,cm$^{-3}$\,yr$^{-1}$ at ages of tens of years (\S~\ref{sub-sec:SNhomog}). A clumpy supernova ejecta may produce the observed DM evolution of FRB\,121102, provided that a substantial fraction of the ejecta mass resides in compact structures of small, $\sim10^{-2.5}$, fractional size  (\S~\ref{sub-sec:SN clumpy}). This is possible in principle, but it is not the generic outcome of current supernova-ejecta models and introduces both geometric and temporal coincidences.
    \item \textit{Confining shell composition.} If the confining plasma is H-dominated and the PRS is efficient in producing ionizing radiation, photo-ionization may dominate ionization by the propagating shock wave, thus suppressing the time variation of the DM due to shock propagation. For a heavier, e.g., C/O dominated, composition, photo-ionization is likely sub-dominant. The observation of a large DM variation therefore disfavors a hydrogen-dominated confining plasma, unless the PRS is an inefficient producer of ionizing radiation.
    \item \textit{Progenitor scenarios.} Among the progenitor channels considered (see Table~\ref{tab:scenarios}), a double WD merger followed by delayed neutron star formation is the most economical interpretation. It naturally separates the epoch of slow mass ejection from the formation of a rapidly rotating neutron star, and predicts a hydrogen-poor, metal-rich confining shell. The main uncertainty is quantitative: existing calculations do not robustly demonstrate the ejection of the required $\sim0.3$--$3\,{\rm M}_\odot$ at tens to hundreds of km\,s$^{-1}$. A common-envelope event followed by a delayed compact-object--core merger provides another possible delay mechanism, but the post-common-envelope in-spiral time and the longevity of the resulting engine are poorly constrained. 
\end{itemize}

Observations of FRB 121102 also show a very large and decreasing RM. The RM declined from $\approx1.5\times10^5$ to $\approx10^5$\,rad\,m$^{-2}$ between 2017 and 2019, with superposed short-timescale variations of $\sim10^3$\,rad\,m$^{-2}/$week \citep{2018Michilli-121102-RM,2021Hilmarsson-121102-RM,2022-Plavin-121102-RM}. It is important to note that the shocked confining plasma may naturally account for the large RM and its variation. The shock driven by the PRS into the confining plasma is collisionless, and thus a significant fraction of the post-shock energy density is expected to be carried by magnetic fields. Using the post-shock pressure given by Eq.~(\ref{eq:p}), the magnetic field carried by the shocked plasma is $B\approx10\epsilon^{1/2}_{B,-2}$\,mG, where $\epsilon_B=10^{-2}\epsilon_{B,-2}$ is the fraction of post-shock thermal energy carried by the magnetic field. The resulting RM is $\approx 0.8 (1/2)(B/1\,\mu{\rm G})(\Sigma/(1\,{\rm pc/cm^3}))\sqrt{l/\Delta R}\approx4\times10^5\epsilon^{1/2}_{B,-2}\Sigma_2\sqrt{l/\Delta R} \,{\rm rad\,m}^{2}$ where $l$ is the field coherence length and $\Delta R\sim R/10$ is the width of the shocked plasma shell. The shocked plasma may naturally account for the observed magnitude of the RM, provided that the field coherence length is not significantly smaller than a few percent of the shocked plasma width. The temporal variability of the RM depends on the evolution of the types and spectra of modes that carry electromagnetic energy far downstream. The study of this evolution is beyond the scope of the current paper. However, we note that 10's of percent variations over several years may be naturally expected, as large variations are to be expected on a time scale $\Delta R/v_s\sim 10\,$yr.

The model makes several near-term predictions. The secular DM should continue to decrease after the breakout, with a gradually decreasing absolute slope as the shell expands. A renewed rise would indicate either interaction with another dense shell or clump, or the late expansion of a previously shocked converging clump. The PRS flux and characteristic synchrotron frequency should evolve on a decade time scale as the relativistic plasma expands and as the engine power changes.

Finally, FRB\,121102 need not be representative of all repeaters. The analysis developed here applies specifically to systems in which a compact relativistic nebula is dynamically confined by a massive local plasma. Long-term DM monitoring of other PRS-associated repeaters will determine whether the rise--breakout--decline sequence is common, and whether a single progenitor channel can account for this subset of the FRB population.

\begin{acknowledgments}
The research of E.W. is partially supported by ISF, Minerva, and Segre grants. E.O.O. is grateful for the support of
grants from the 
Willner Family Leadership Institute,
Andr\'e Deloro Institute,
Benoziyo center for astrophysics,
the Orion center for ground-based astronomy,
Paul and Tina Gardner,
The Norman E Alexander Family M Foundation ULTRASAT Data Center Fund,
Israel Science Foundation,
Israeli Ministry of Science,
Minerva,
BSF, NSF-BSF,
Israel Council for Higher Education (VATAT),
Sagol Weizmann-MIT,
the Institute for Environmental Sustainability (IES) at the Weizmann Institute of Science and by Dr. Veronika A. Rabl, and the
Rosa and Emilio Segr\`e Research Award.  DK is supported by a research grant from The Abramson Family Center for Young Scientists, an ISF grant, the Minerva Stiftung, and the Pazi Foundation.
 
\end{acknowledgments}

\appendix

\begin{deluxetable*}{llccclll}
\tablecaption{FRB-associated persistent radio sources}
\tablehead{
\colhead{FRB} &
\colhead{R.A. (J2000)} &
\colhead{Decl. (J2000)} &
\colhead{DM} &
\colhead{$z$} &
\colhead{$L_\nu$} &
\colhead{$(\nu)$} &
\colhead{$R$}
\\
\colhead{} &
\colhead{} &
\colhead{} &
\colhead{$({\rm pc\,cm^{-3}})$} &
\colhead{} &
\colhead{$({\rm erg\,s^{-1}\,Hz^{-1}})$} &
\colhead{} &
\colhead{}
}
\startdata
FRB 121102A &
05:31:58.70 & +33:08:52.5 &
$\simeq 558$ &
0.19273 &
$\sim 2$--$3\times10^{29}$ $(1.4$--$1.7\,{\rm GHz})$ & $\nu^0$ & $<0.3\,$pc
\\
FRB 20190520B   &
16:02:04.27 & $-11$:17:17.3 &
$\simeq 1205$ &
0.241 &
$(3.0\pm0.5)\times10^{29}$ $(1.7\,{\rm GHz})$  &
$
\nu^{-0.41\pm0.04}$ & $<9$\,pc
\\
FRB 20201124A   &
05:08:03.51 & +26:03:38.5 &
$\simeq 413$ &
0.098 &
$1.2\times10^{29}$ $(1.6\,{\rm GHz})$ 
$4.9\times10^{27}$ $(15\,{\rm GHz})$ & &
\\
FRB 20240114A  &
21:27:39.84 & +04:19:45.7 &
$527.65\pm0.01$ &
$\simeq 0.13$ &
$\simeq 2.2\times10^{28}$ $(5\,{\rm GHz})$ & &
\\
FRB 20190417A  &
19:39:05.90 & +59:19:36.8 &
$\simeq 1379$ &
0.12817 &
$\simeq 8\times10^{28}$ $(1.4\,{\rm GHz})$ & $\nu^{-0.19\pm0.29}$ & $<23$\,pc
\enddata
\tablecomments{
FRB\,121102A data are discussed in \S~\ref{sub-sec:PaperIconst}. Luminosities are spectral radio luminosities and are frequency-dependent. Data for FRB\,20190520B are taken from \cite{Niu+2022Nature_FRB20190520B_Detection_PRS}
and \cite{Bhandari+2023ApJ_FRB20190520B_PRS_size}, 
for FRB\,20201124A
from \cite{Ravi+2022MNRAS_FRB20201124A_Host_PRS} and \cite{Bruni+2024Nature_FRB20201124A_PRS} (The 201124A 1.6~GHz flux is attributed to a star-forming region),
for FRB\,20240114A from
\citet{2025Bruni-20240114A},
and for FRB\,20190417A
from \cite{Bruni+2026_FRB20190417A_PRS} and \cite{2026Moroianu-PRS}.
}
\label{tab:PRSs}
\end{deluxetable*}

\section{A. Including expansion and radial density dependence}
\label{app:detailed}

Consider a shock driven by a time-independent, high-pressure into a spherically symmetric homologously expanding shell with a power-law density profile, $n_c\propto r^{-\alpha_\rho}t^{-(3-\alpha_\rho)}$. The time-dependent DM is given by
\begin{equation}\label{eq:DM_t}
  DM=\frac{N_s}{4\pi r_s^2}=\frac{1}{3-\alpha_\rho}n_c(t,r=r_s)r_s,
\end{equation}
where $N_s=4\pi r_s^3 n_c/(3-\alpha_\rho)$ is the number of electrons in the shocked plasma and $r_s$ is the shock radius (the shocked plasma is confined to a thin, $<r_s/12$ shell behind the shock (\S~\ref{sub-sec:DMvar}); note that the density of a shocked fluid element is time independent since the pressure is time independent and the flow is adiabatic). The time derivative of the DM is given by
\begin{eqnarray}
    \label{eq:DMdot}
    \dot{DM} &=& \frac{\dot{N}_s}{4\pi r_s^2}-\frac{2N_s\dot{r}_s}{4\pi r_s^3}=v_s n_c-\frac{2}{3-\alpha_\rho}n_c(v_0+v_s)
    \nonumber \\ &=&\frac{1-\alpha_\rho-2v_0/v_s}{3-\alpha_\rho} n_cv_s,    
\end{eqnarray}
where $v_0$ is the pre-shocked fluid velocity (at $r_s)$ and $\dot{N_s}=4\pi r_s^2 n_c v_s$ since $v_s$ is the shock velocity with respect to the fluid. When the shock approaches the outer shell radius, $R$, at the peak DM time, $t_p$,
\begin{equation}\label{eq:DMdot-}
  \dot{DM} = (1-\alpha_\rho-2\alpha_v)\frac{v_s}{R} \Sigma
  \quad {\rm at}\, t<t_p,
\end{equation}
where $\alpha_v\equiv v_0(t_p,R)/v_s$ and $\Sigma= N_s(r_s=R)/(4\pi R^2)$. The DM increase rate is suppressed by the decline of density with radius, $\alpha_\rho$, and by its decline with time due to (homologous) expansion, $\alpha_v$. The latter depends on the ratio between expansion rate, $v_0/R$, and the shock propagation rate $v_s/R$.

After shock crossing of the shell, $DM=\Sigma R^2/(R+(v_s+v_0)t)^2$ implying
\begin{equation}\label{eq:DMdot+}
  \dot{DM} = -2\frac{(v_s+v_0) R^2}{(R+(v_s+v_0)t)^3}\Sigma 
  \approx -2(1+v_0/v_s)\frac{v_s}{R}\Sigma
\end{equation}
at $t>t_p$. This implies
\begin{equation}\label{eq:DMrat_app}
 \frac{\dot{DM}_+}{\dot{DM}_-}=-2\frac{(1+v_0/v_s)}{(1-\alpha_\rho-2v_0/v_s)}.
\end{equation}

\section{B. Hydrodynamics of shocks driven by energy injection at the center of a homologously expanding ejecta: Delayed onset of energy injection}
\label{app:numeric1}

Consider a cold plasma sphere with uniform density expanding homologously, with radius $R_0$, density $\rho_0$, and velocity $v=(r/R_0)v_0$ at $t=0$. At $t>0$ energy is injected at the center, $r=0$, at a time-independent rate, $\dot{E}$, in the form of a highly relativistic plasma. A shock driven by the relativistic plasma pressure propagates through the cold shell. Since the speed of sound in the relativistic plasma is close to $c$, we approximate the pressure within the expanding inner sphere of the relativistic plasma as uniform. Denoting the radius of the inner sphere by $r_{\rm in}$ and the relativistic plasma energy by $E_r$, we have
\begin{equation}
    \label{eq:rel_sphere}
    p_r=\frac{E_r}{4\pi r_{\rm in}^3},\quad \dot{E}_r=\dot{E}-\frac{\dot{r}_{\rm in}}{r_{\rm in}} E_r.
\end{equation}

The initial kinetic energy of the cold plasma is
\begin{equation}
    \label{eq:Ek_0}
    E_k=\frac{2\pi}{5}\rho_0R_0^3v_0^2.
\end{equation}
Measuring distance in units of $R_0$, time in units of $t_0\equiv R_0/v_0$, and mass in units of $\rho_0R_0^3$, the flow is completely defined by a single dimensionless parameter,
\begin{equation}
    \label{eq:eta_app}
    \eta\equiv\frac{\dot{E}t_0}{E_k}=\frac{5}{2\pi}\frac{\dot{E}}{\rho_0R_0^2v_0^3}.
\end{equation}

To obtain a shock velocity at breakout, $v_s \gg v_0$, $\eta\gg1$ is required. In this limit, we may obtain an approximate solution by neglecting the initial velocity of the cold plasma. The post shock energy is composed of the thermal energy,
\begin{equation}
    \label{eq:E_ther}
    E_{\rm th} \simeq 4\pi r_s^3 p_r = 4\pi r_s^3 \alpha \frac{2}{\gamma+1} \rho v_s^2,
\end{equation}
where $\gamma$ is the adiabatic index of the shocked plasma and $\alpha\lesssim1$. The kinetic energy of the shocked plasma is
\begin{equation}
    \label{eq:E_s}
    E_s=\frac{2\pi}{3}\rho r_s^3 \left(\frac{2}{\gamma+1}v_s\right)^2.
\end{equation}
Energy conservation,
\begin{equation}
    \label{eq:E_cons}
    \dot{E}t=4\pi r_s^3\rho v_s^2 \frac{2 }{\gamma+1}\left(\alpha +
    \frac{1}{3(\gamma+1)}\right),
\end{equation}
implies $r_s=A t^{3/5}$ and $v_s=(3/5)r_s/t$, thus
\begin{equation}
    \label{eq:E_cons_const}
    \dot{E}=\frac{36\pi}{25} \frac{2}{\gamma+1}\left(\alpha +
    \frac{1}{3(\gamma+1)}\right) A^5\rho_0.
\end{equation}

We may obtain another equation from the conservation of momentum
\begin{eqnarray}
    \label{eq:P_cons}
    4\pi r_s^2 \alpha \frac{2}{\gamma+1} \rho v_s^2 &=& 
    \frac{d}{dt}\left(M\frac{2}{\gamma+1} v_s\right)
    \nonumber \\ &=&\frac{2}{\gamma+1}\frac{d}{dt}\left(\frac{4\pi}{3}\rho r_s^3 v_s\right)
    = \frac{2}{\gamma+1}\frac{7}{5}\frac{4\pi}{3t}\rho r_s^3 v_s
    \nonumber \\ &=& \frac{2}{\gamma+1}\frac{7}{3}\frac{4\pi}{3}\rho r_s^2 v_s^2,
\end{eqnarray}
yielding
\begin{equation}
    \label{eq:alpha_A}
     \alpha  = \frac{7}{9}, \quad A = 0.80 \left(\frac{\dot{E}}{\rho_0}\right)^{1/5},
     \quad r_s =  0.80 \left(\frac{\dot{E}}{\rho_0}t^3\right)^{1/5}.
\end{equation}
This gives the shock breakout time, $t_b$,
\begin{equation}
    \label{eq_t_b}
    \frac{t_b}{t_0}=0.80^{-5/3} \left(\frac{R_0^5\rho_0}{t_0^3\dot{E}}\right)^{1/3}=
    0.80^{-5/3} \left(\frac{5}{2\pi\eta}\right)^{1/3}=1.34\eta^{-1/3},
\end{equation}
and shock velocity at breakout,
\begin{equation}
    \label{eq:v_b}
    \frac{v_{s,b}}{v_0}=\frac{(3/5)R_0/t_b}{R_0/t_0}=0.45\eta^{1/3}.
\end{equation}

Solving the hydro equations numerically for $\eta=10^3,10^4$ gives $t_b/t_0=0.14,0.062$ compared to the analytic values $0.13,0.062$ given by Eq.~(\ref{eq_t_b}), and $v_{s,b}/v_0=5.4,10.9$ compared to the analytic values $4.5,9.7$ given by Eq.~(\ref{eq:v_b}). Including in the analytic solution the shock acceleration due to expansion, which enhances the velocity by a factor $(1+t_b/t_0)^{3/2}$, yields $v_{s,b}/v_0=5.4$ and $10.6$. The analytic and numerical results agree.

\section{C. Hydrodynamics of shocks driven by energy injection at the center of a homologously expanding ejecta: Contemporaneous explosion and onset of energy injection}
\label{app:numeric2}

Consider a cold plasma sphere with uniform density expanding homologously, with mass $M$ and maximal velocity $v_0$. We set $t=0$ to be the time of expansion onset (so that the outer radius is given by $R=v_0t$), and assume that energy is injected for $t>0$ at the center, $r=0$, with a time-independent rate, $\dot{E}$, in the form of a highly relativistic plasma. A shock driven by the relativistic plasma pressure propagates through the cold expanding shell. 

Motivated by the fact that the injected energy, $\dot{E}t$, is likely to affect the flow up to the radius out to which the initial kinetic energy of the cold plasma is similar to $\dot{E}t$, we look for a solution to the propagating shock radius, $r_s(t)$, of the form
\begin{equation}
    \label{eq:rs_R_no-delay}
    \frac{r_s}{R} = \xi\left(\frac{\dot{E}t}{E_k}\right)^{1/5},
\end{equation}
where $E_k=0.3Mv_0^2$ is the initial kinetic energy of the cold plasma (Note that the fraction of the kinetic energy contained by the plasma out to $r$ is $(r/R)^5$). The shock velocity with respect to the fluid, $v_s=\dot{r}_s-r_s/t$ where $v=r_s/t$ is the fluid velocity at the shock's position, is in this case
\begin{equation}
    \label{eq:vs_no-delay}
    v_s=\dot{r}_s-\frac{r_s}{t} = \frac{1}{5}\frac{r_s}{t}=\frac{1}{5}v.
\end{equation}

Let us consider now the conservation of energy. The kinetic energy of the shocked plasma is $0.5M(r_s)(v+2v_s/(\gamma+1))^2$, where $M(r_s)$ is the mass contained out to $r_s$, $2v_s/(\gamma+1)$ is the post-shock fluid velocity in the upstream fluid frame, and $\gamma=5/3$ is the adiabatic index of the non-relativistic gas. Since the pre-shock kinetic energy is $0.3M(r_s)v^2$, the kinetic energy imparted by the shock to the cold plasma is
\begin{equation}
    \label{eq:Ek_no-delay}
    \Delta E_k = 2x_s\left(1+ x_s + 0.1x_s^{-1}\right)M(r_s)v^2,
\end{equation}
with
\begin{equation}
    x_s\equiv \frac{v_s}{(\gamma+1)v}.
\end{equation}
The post-shock thermal energy density is $2\rho v_s^2/(\gamma+1)$, where $\rho$ is the upstream density. Approximating the pressure within the relativistic plasma sphere as nearly uniform, noting that most of the volume up to $r_s$ is occupied by the relativistic plasma for which the pressure is $1/3$ the energy density (the shocked confining plasma is compressed to a narrow shell), and assuming that the relativistic plasma pressure is a time independent factor $\alpha\gtrsim1$ larger than the post shock pressure (note that the shocked plasma is accelerating), the thermal energy is given by
\begin{equation}
    \label{eq:Eth_no-delay}
    E_{th} = 4\pi r_s^3\frac{2\alpha}{(\gamma+1)}\rho v_s^2 = \frac{6\alpha v_s^2}{(\gamma+1)v^2}M(r_s)v^2\,.
\end{equation}
Energy conservation, therefore, implies
\begin{eqnarray}
    \label{eq:Econ_no-delay}
    \dot{E}t = 2x_s\left(1 + x_s + 3 \alpha\frac{v_s}{v}+\frac{0.1}{x_s}\right) \left(\frac{r_s}{R}\right)^5 \frac{E_k}{0.3}\, . 
\end{eqnarray}
Our assumed solution, Eq.~(\ref{eq:rs_R_no-delay}), which implies a time independent $v_s/v=1/5$ and $r_s/R \propto t^{1/5}$, allows satisfying energy conservation at all $t$ for
\begin{equation}
    \label{eq:xi_alpha}
    \xi^5 = 0.3 \left[2x_s\left(1 + x_s + 3 \alpha\frac{v_s}{v}+\frac{0.1}{x_s}\right)\right]^{-1}.
\end{equation}

$\alpha$ may be obtained from the conservation of momentum of the narrow shocked plasma shell,
\begin{eqnarray}
    \label{eq:P_cons-no-delay}
    4\pi r_s^2 \left(\frac{2\alpha }{\gamma+1}+\frac{v}{v_s}\right) \rho v_s^2 &=& 
    \frac{d}{dt}\left[\left(1+\frac{2v_s}{(\gamma+1)v}\right)v M(r_s)\right]
    \nonumber \\ &=& \left(1+2x_s\right)Mv_0\frac{d}{dt}\left(\frac{r_s}{R}\right)^4
    \nonumber \\ &=& \left(1+2x_s\right)Mv_0\frac{4}{5t}\left(\frac{r_s}{R}\right)^4
    \nonumber \\ &=& \left(1+2x_s\right)\frac{16\pi}{15}r_s^2\rho v^2,
\end{eqnarray}
where the first term on the l.h.s. represents the momentum change of the shocked shell due to the force applied by pressure, and the second accounts for the contribution of the pre-shocked plasma momentum. This implies
\begin{eqnarray}
    \label{eq:alpha-no-delay}
  \alpha= \left(\frac{(\gamma+1)v}{2v_s}+1\right)\frac{4}{15} \frac{v}{v_s} - \frac{(\gamma+1)v}{2v_s} = 3.6,
\end{eqnarray}
and
\begin{equation}
    \label{eq:xi}
    \xi = 0.85.
\end{equation}
Eqs.~(\ref{eq:rs_R_no-delay}) and~(\ref{eq:xi}) provide an excellent approximation to the results of numeric calculations.

\bibliographystyle{hapj}

\end{document}